\documentclass[12pt]{article}
\usepackage[final]{changes}
\makeatletter
\colorlet{Changes@Color}{teal}
\makeatother

\usepackage[onehalfspacing]{setspace}

\usepackage{natbib}
\usepackage{graphicx}
\usepackage{caption}
\usepackage{subcaption}
\usepackage{amsmath,amssymb,amsfonts,mathtools,placeins}
\usepackage{color}
\usepackage{relsize}
\usepackage{pgfplots}
\usepackage{enumitem}
\pgfplotsset{compat=1.12}

\usepackage{authblk}

\newtheorem{prop}{Proposition}

\newtheorem{assume}{Assumption}
\newtheorem{example}{Example}

\usepackage{chngcntr}
\usepackage{apptools}
\AtAppendix{\counterwithin{lemma}{section}}
\AtAppendix{\counterwithin{prop}{section}}

\title{Debunking Rumors in Networks\thanks{We thank Leonie Baumann, Francis Bloch, Ugo Bolletta and Tom\`{a}s Rodgriguez-Barraquer, and (seminars) participants at CTN 2016, SAET 2016, BiNoMa 2017, LAGV 2017, the $4^{th}$ Conference on Network Science and Economics, the 2019 VERA workshop, Auton\`{o}ma Barcelona, Navarra, Universit\'{e} libre de Bruxelles and Virginia Tech for valuable comments. Merlino gratefully acknowledges funding from the CNRS and the Research Foundation Flanders through grant G029621N. Pin gratefully acknowledges funding from the Italian Ministry of Education Progetti di Rilevante
Interesse Nazionale (PRIN) grant 2017ELHNNJ and from Regione Toscana grant Spin.Ge.Vac.S..
Tabasso gratefully acknowledges funding from the European Union's Horizon 2020 research and innovation programme under the Marie Sk\l{}odowska-Curie grant agreement No. 793769. \protect\includegraphics[height=.8\baselineskip]{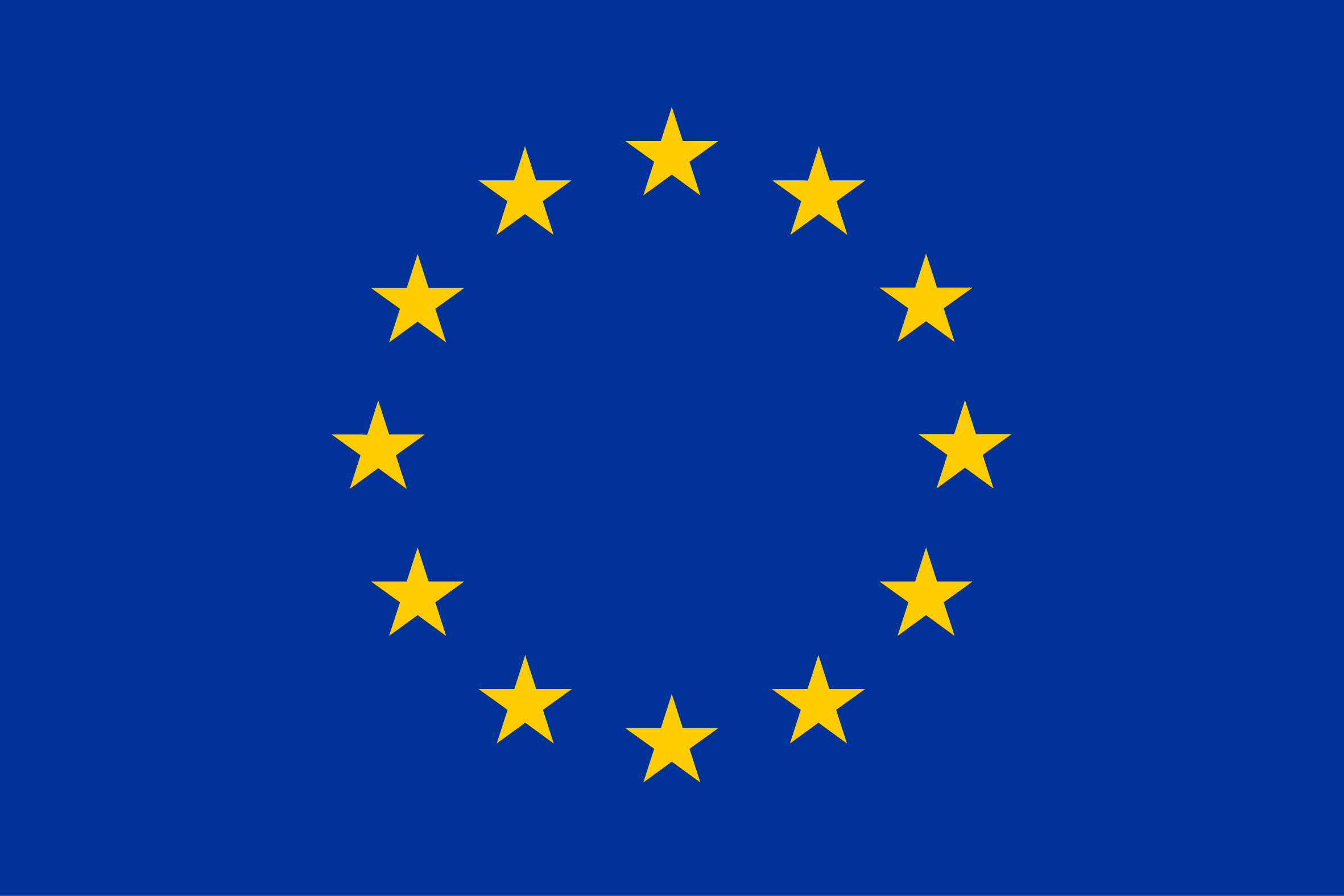} 
}}

\author[a,b]{Luca P. Merlino\thanks{Corresponding author: LucaPaolo.Merlino@uantwerpen.be}}
\author[c,d]{Paolo Pin}
\author[e,f]{Nicole Tabasso}
{
\affil[a]{\footnotesize Department of Economics University of Antwerp, Antwerp, Belgium 
}
\affil[b]{ECARES, Universit\'{e} libre de Bruxelles, Belgium}
\affil[c]{Department of Economics and Statistics, Universit\`a di Siena, Italy 
}
\affil[d]{BIDSA, Universit\`a Bocconi, Milan, Italy}
\affil[e]{Department of Economics, Universit\`a Ca' Foscari, Venice, Italy 
}
\affil[f]{School of Economics, University of Surrey, UK}
}

\date{}

\begin{document}
\maketitle

\vspace{-1cm}
\begin{abstract}
We study the diffusion of a true and a false message (the rumor) in a social network. Upon hearing a message, individuals may believe it, disbelieve it, or debunk it through costly verification. Whenever the truth survives in steady state, so does the rumor. \added{Communication intensity in itself is irrelevant for relative rumor prevalence, and the effect of homophily depends on the exact verification process and equilibrium verification rates.} Our model highlights that successful policies in the fight against rumors increase individuals’ incentives to verify.
\end{abstract}

\thispagestyle{empty}
\setcounter{page}{0}

\newpage

\section{Introduction}

Information often diffuses via communication with family, friends or acquaintances. However, people transmit not only correct information, but also rumors, that is, false or imprecise information. The virality of these rumors shapes public debates, often involving significant personal and social costs.\footnote{\added{For example, believing conspiracy theories that deny the link between HIV and AIDS is associated with a less consistent use of condoms in the US \citep{Bogart}.}}

The increased reliance on online social media for news consumption and communication plays an important role in the diffusion of rumors. \added{This has been documented in different contexts, such as fake news during the 2016 US Presidential election campaign \citep{Allcott}, the dangers of childhood vaccinations \citep{Needle} and the origins of COVID-19 \citep{mian2020}.}
\added{A main concern is the degree of homophily in online social media, inducing ``echo chambers'' in which people are over-proportionally exposed to one particular opinion.}\footnote{\added{While online patterns of news consumption are no more segregated than offline ones \citep{Gentzkow,Halberstam}, online social networks appear to be extremely homophilous (see, e.g., \citealp{zollo2017}) and lead to more segregated communication patterns \citep{Halberstam}.}} 

This concern has prompted demands on policy makers, news providers and online social networks to counter the diffusion of rumors \citep{WarRoom}. \added{For example, misinformation relating to the COVID-19 pandemic has caused responses from various platforms, from showing warnings on messages coming from debatable sources to providing links to verified and authoritative ones \citep{Marr}.}


\added{This paper proposes a tractable model to understand how the diffusion of rumors is affected by endogenous debunking and changes in homophily, as well as to derive policy implications. Crucially, our model allows for the persistence of two distinct messages about the state of the world (the truth and the rumor) when messages can be verified.}

In our model, individuals receive a message from their \added{social} contacts. \added{Society is partitioned into two groups, where individuals of each group have a prior $y>0.5$ that the true state of the world is either $0$ or $1$, respectively. We say that individuals are biased towards a certain state.}\footnote{It has been documented that people who tend to believe in fake news can clearly be identified in society and in online social communities (e.g., \citealt{zollo2017}; \citealt{samantray2019credibility}). A micro-foundation for this segregation is in \cite{bolletta2019polarization}.} \added{An individual can exert verification effort, which reveals the true state with some probability. This effort is determined by their posterior belief about the state of the world upon receiving a message, calculated using Bayes rule.}

Upon successful verification, individuals become aware of the true state of the world and accept it. If verification is unsuccessful, individuals' \added{opinion matches the state towards which they are biased.} Individuals communicate their opinion to their neighbors in a network, but they cannot credibly reveal their prior, nor whether they verified. The network is characterized by a degree of homophily, which is the probability with which a neighbor has the same prior as oneself.

In this model, \added{information prevalence converges to a steady state in which} the rumor always survives alongside the truth: \added{as verification is costly, it is not perfect, so the rumor propagates.}
\added{The truth to rumor ratio depends on the level of verification and the degree of homophily in the network. Indeed, an increase in communication rates increases both rumor and truth in equal proportion, such that the ratio remains constant. The degree of homophily instead creates trade-offs. As it increases, individuals receive relatively more messages that confirm their prior. As these are verified less than opposing messages, homophily creates echo chambers and benefits the rumor. However, higher homophily also makes messages reinforcing one's prior less informative, and the opposite for messages going against one's prior. This incentivizes verification, which reduces the diffusion of the rumor. Which effect dominates depends on how verification efforts translate into verification success. We derive general conditions on the verification function that determine the impact of changes in homophily. We also discuss an example that highlights} the role of the verification technology in shaping how homophily affects the quality of information.

\added{Our model highlights that} any policy which amounts to a one-time injection of truthful information, such as a time-limited information campaign, is ineffective in reducing the ratio of rumor to truth in the long run. Policies should rather incentivize individuals to verify information, thereby increasing the truth to rumor ratio. For example, strategies to combat fake news that focus on providing links to sources of verification are likely more effective than simply flagging posts as being disputed.

Our results are robust to the introduction of partisans who \added{always hold the same opinion}. Indeed, these individuals \added{provide} higher incentives to verify \added{to the rest of the population}, so as to completely offset their presence. Ignoring endogenous debunking would lead to very different conclusions.

Our paper contributes to the literature using game theory to study information diffusion in networks. \added{This literature has two main strands. In the first, agents are Bayesian, e.g., \cite{Hagenbach-Koessler} and \cite{Galeotti-Christian}.} More related to our paper, \cite{Bloch-Rumors} find that, when there are partisans who diffuse false information, other agents block messages coming from parts of the network with many partisans. \added{\cite{kr20} study how strategic diffusion affect media quality. In our paper, there is no strategic motive to transmit messages. However, debunking introduces a different kind of interaction: the rumor's prevalence affects the incentives of individuals to verify.}

\added{The other strand of literature considers non-Bayesian agents who learn from their neighbors. Starting from the seminal contribution of \cite{deGroot}, in these models agents observe the beliefs of their neighbors and use them to update theirs using some specific rule \citep{banerjee2004word,Golub-Jackson}.}

\added{Recently, some papers have studied the behavior of Bayesian agents when the underlying information structure is misspecified, e.g., \cite{molavi2018} and \cite{banerjee2020}. In a similar spirit, in our model agents act Bayesian when they first hear a message, but they disregard additional information once they have formed an opinion. Possible interpretations are that an opinion translates into a once-for-all decision or individuals are unwilling to change their opinion. Such behavior is consistent with information avoidance, inattention, or a biased interpretation of additional information \citep{golman2017information}.}

\added{In our model, individuals do not observe the diffusion of messages, but they derive them by the properties of the diffusion process assuming that their prior is correct. Since the prior might be wrong, individuals might hold wrong beliefs in the long run, as in, e.g., \cite{compte2004}.}

\added{We employ a} $SIS$ framework, a class of models introduced to study the diffusion of viruses.\footnote{$SIS$ stands for Susceptible-Infected-Susceptible as infected individuals return to the susceptible class on recovery as the disease confers no immunity against reinfection.} Following the seminal work of \added{\cite{banerjee1993economics} and} \cite{kremer1996}, some papers have studied how strategic decisions on protection affect the diffusion of a disease \citep{toxvaerd2014,Goyal-Vigier,toxvaerd2019,bizzarri2020segregating}. In particular, \cite{Galeotti-Rogers-AEJ} study the effect of homophily on strategic immunization. \added{In these papers, there is a unique infectious state, whose magnitude is affected by immunization. We instead focus on the relative magnitude of truth and rumor within the overall prevalence of information, which implies different strategic considerations. In particular, while in the above papers protection is a local public good \citep{KM}, this is not true for discordant messages in our framework.}

\added{This feature also distinguishes our paper from the recent contributions on the role of costly search on social learning \citep{ali2018herding,mueller2016}. As in our paper, learning is not complete (i.e., beliefs do not converge to the truth) precisely because search (here, verification) is costly.}

\added{We study the diffusion of two messages as in \cite{clz19} and \cite{Tabasso1}. Contrarily to these paper, here we consider contradictory pieces of information that may be disbelieved.} More broadly, our paper relates to a recent literature on opinion dynamics on random graphs \citep{akbarpour2018,sadler2020}. 
In those papers, agents either adopt or not without the option of external verification, and non-adoption does not create any externality. In our model, one's decision depends instead on the level of (non-)verification in the economy.

We believe we are the first to study the strategic decision of individuals to verify what they hear when there are several messages diffusing in a network.

The paper proceeds as follows. Section \ref{model} introduces \added{and discusses} the model. Section \ref{main} presents the main analysis. Section \ref{policies} discusses the policy implications. Section \ref{sec:partisan} presents the model with partisans. Section \ref{conclusions} concludes. All proofs \added{and computations for the examples} are in the Appendix.


\section{The Model}\label{model}

\added{In this section, we formally present the model. The timeline is as follows. An individual $i$ who hears a message at time $t$ chooses how much effort to exert in verifying it. They then form an opinion of the true state of the world. While alive, they communicate a message in line with their opinion to their social contacts at a fixed rate.}

\added{In the following, we first describe the $SIS$ diffusion process; then we derive the differential equations that govern the evolution of truth and rumor, given verification rates. Finally, we study how individuals chose these rates. We end the section with a discussion of the main assumptions of our model.}

\bigskip

\noindent \added{\textbf{Diffusion Process.}}  There is an infinite population of mass $1$ of individuals, indexed by $i$, represented as nodes on a network. Time is continuous, indexed by $t$. There exist two \added{verifiable} messages $m\in\{0,1\}$ that individuals diffuse via word of mouth. These messages pertain to the state of the world, $\Phi\in\{0,1\}$. Without loss of generality, we denote $\Phi=0$ as the true state of the world, \textit{ex-ante} unknown to the individuals. We refer accordingly to $m=1$ as the ``rumor''. When individual $i$ communicates message $m$ to individual $j$, this reveals to $j$ the set of values that $\Phi$ may take.

Society is partitioned into two groups of equal size, denoted by $b=\{0,1\}$, where individuals of each group have a prior $y>.5$ that the true state of the world is either $0$ or $1$, respectively.

Each individual has $k$ meetings at each time $t$. A proportion $\beta\in[0,1]$ of these is with individuals of the same group, while the remaining interactions are with individuals of the other group. The group one belongs to is not observable, but $\beta$ is common knowledge. A meeting between two individuals is described by a link. The associated network is realized every period.

\added{The diffusion process of information is a $SIS$ model.} Individuals may be in one of two states: either they are unaware of the debate about the state of the world, in which case they are in state $S$ (\textit{Susceptible}), or they may hold an opinion about its value, in which case they are in state $I$ (\textit{Informed}).

\begin{figure}[htbp!]
 \centering
 \includegraphics[width=10.5cm]{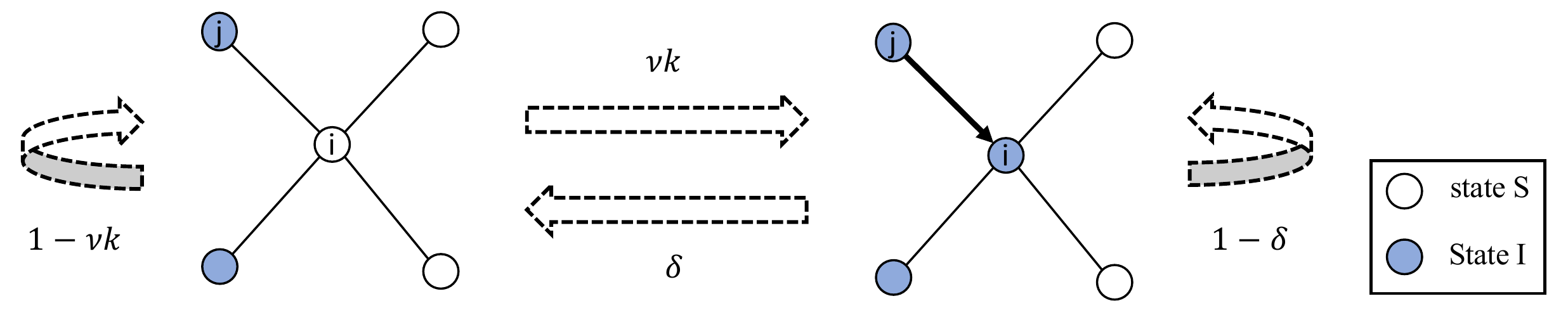}
 \caption{The transition dynamics of player $i$ for $k=4$.}
 \label{fig:SIS}
\end{figure}

An individual in state $S$ transitions into state $I$ by hearing message $m$ during a meeting, in which communication occurs at rate $\nu$. \added{We assume that $\nu$ is sufficiently small that the chance of receiving multiple messages at the $k$ simultaneous meetings is zero, so that information transmits at rate $k\nu$. With the complementary probability, an individual in $S$ stays in $S$.} Individuals in state $I$ die at rate $\delta$ and are replaced with individuals of the same type in state $S$. \added{Figure \ref{fig:SIS} depicts the transmission dynamics for an individual $i$, who becomes informed after receiving a message from $j$.}

\bigskip

\begin{figure}[ht!]
 \centering
 \includegraphics[width=12cm]{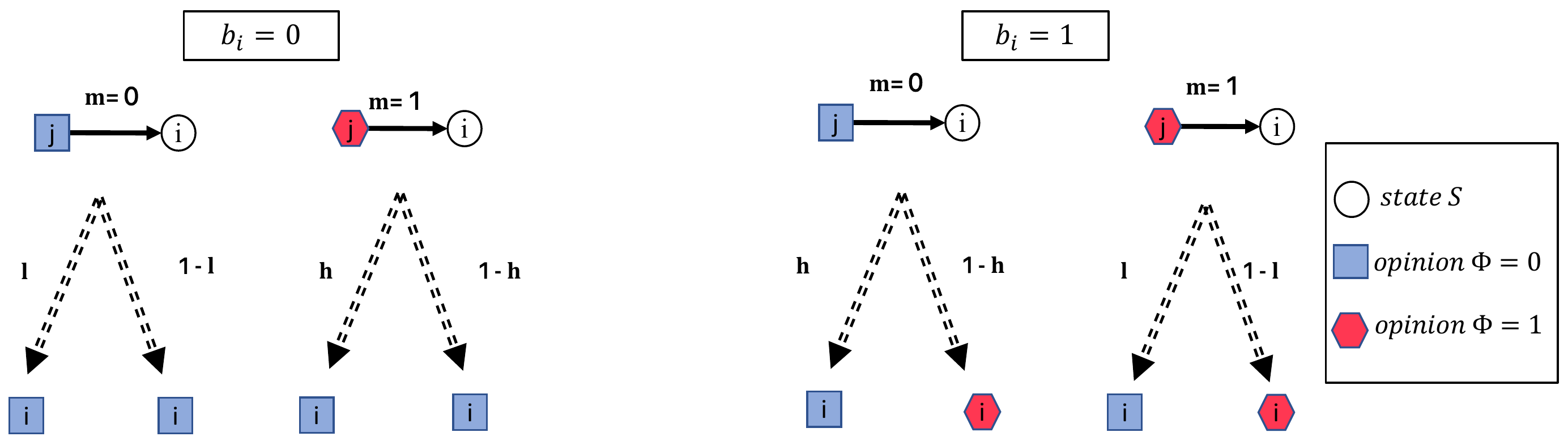}
 \caption{A summary of the potential opinions $i$ may hold, depending on her type, the message received, and verification success.}
 \label{fig:opinions}
\end{figure}

\noindent \added{\textbf{Information Prevalence.} Individuals who hear $m$ choose how much effort to exert in verification. In Section \ref{endogenous-v}, we show that this effort depends on the type of message an individual receives. We define $l_t$ as the rate of verification of messages in line with one's bias (i.e., $m=b$), and $h_t$ the verification rate of messages which go against it (i.e., $m\neq b$). Successful verification implies that} an individual $i$ knows that $\Phi=0$ for sure\added{; hence, $i$ will hold this opinion}. With the complement probability, the result of the verification process is inconclusive. \added{In that case, an individual of type $b$ holds an opinion in line with their bias (i.e., $\Phi=b$). We derive in Proposition \ref{Existence} when this is the optimal behavior. Figure \ref{fig:opinions} depicts how individuals of type $0$ and $1$ form their opinion.}

Denote by $\rho^b_{m,t}(\ell_t,h_t)$ the proportion of type $b$ individuals that hold opinion in line with $m$ at time $t$.\footnote{Since $k=k_i$ $\forall i \in N$, the proportion of individuals with degree $k$ who hold opinion in line with $m$ is identical to the overall proportion of individuals who hold that opinion.} Note that type $0$ individuals may only hold opinion $0$, irrespective of verification, i.e., $\rho^0_{1,t}(\ell_t,h_t)=0$. With some abuse of notation, we usually suppress the dependence on $(\ell_t,h_t)$.

The laws of motion governing the transmission of the system then are:
\begin{small}
\begin{align}
\frac{\partial \rho_{0,t}^{0}}{\partial t} &=
\frac{1}{2}(1-\rho^0_{0,t})\nu k \left[\beta\rho^0_{0,t}+(1-\beta)(\rho^1_{0,t}+\rho^1_{1,t})\right] - \frac{1}{2}\rho^0_{0,t}\delta, \label{rho00} \\
\frac{\partial \rho_{0,t}^{1}}{\partial t} &= \frac{1}{2}(1-\rho_{0,t}^1-\rho_{1,t}^1)\nu k 
 \left[\beta \left(h_t \rho_{0,t}^1+\ell_t\rho_{1,t}^1\right) +(1-\beta)h_t\rho_{0,t}^0 \right] 
- \frac{1}{2}\rho_{0,t}^1\delta ,\label{rho01}
\end{align}
\begin{align}
\frac{\partial \rho_{1,t}^1}{\partial t} &= \frac{1}{2} (1-\rho_{0,t}^1-\rho_{1,t}^1)\nu k 
 \left[\beta\left((1-h_t)\rho_{0,t}^1+ \nonumber \right. \right. \\ & \left.\left. \qquad \qquad \qquad \qquad \qquad + 
 (1-\ell_t)\rho_{1,t}^1 \right) +(1-\beta)(1-h_t)\rho_{0,t}^0\right] - \frac{1}{2}\rho_{1,t}^1\delta , \label{rho11}
\end{align}
\end{small}
\noindent These expressions describe the evolution of opinions in line with $m=0$ and $m=1$ within the two groups $b\in\{0,1\}$ \added{in the \textit{mean-field approximation} of the system, whereby information prevalence in an individual's neighborhood is the same as the prevalence in the overall population. For example, \eqref{rho00} describes how $m=0$ evolves in group $b=0$. The first term represents the mass of individuals who start holding opinion $\Phi=0$ at time $t$. Indeed, the proportion of susceptible individuals of type $0$, ($1-\rho^0_{0,t}$), receive a message at rate $\nu k$. The message can come from someone of the $\rho^0_{0,t}$ individuals of their group, who holds opinion $0$ and whom they meet with probability $\beta$. With probability $1-\beta$, they meet someone of group 1, of whom $\rho^1_{0,t}$ transmit message $0$ and $\rho^1_{1,t}$ message $1$.} The second (negative) term, indicates that a proportion $\delta$ of the informed individuals of type $0$, $\rho^0_{0,t}$, die and are replaced by individuals in state $S$. \added{In sum, players of group $0$ always hold an opinion $0$ independently of verification (see the left panel of Figure \ref{fig:opinions}).}

\added{This is not the case for individuals of group $1$. For example, consider how $m=0$ evolves in group $b=1$---equation \eqref{rho01}. When the proportion of susceptible individuals of type $1$, ($1-\rho_{0,t}^1-\rho_{1,t}^1$), receive a message, they need to successfully verify the messages they receive to hold opinion $0$. This happens with probability $\ell_t$ for messages in line of their bias, which they receive from $\rho_{1,t}^1$ individuals of their group, who they meet with probability $\beta$. They verify with probability $h_t$ messages such that $m=0$, which they can receive from $\rho_{0,t}^1$ of their group, who they meet with probability $\beta$, or from people in the other group, $\rho_{0,t}^0$, who they meet with probability $1-\beta$. Again, informed individuals die and are replaced by individuals in state $S$ at rate $\delta$. The interpretation of \eqref{rho11} is equivalent. These transitions are depicted in the right panel of Figure \ref{fig:opinions}.}

Our main objects of interest are the overall prevalence of opinions $0$ and $1$ in the population at time $t$, $\rho_{m,t}$, which are
\begin{eqnarray}
\rho_{0,t} &=& \frac{1}{2} \left(\rho^0_{0,t} +\rho^1_{0,t}\right), \\ 
\rho_{1,t} &=& \frac{1}{2}\rho^1_{1,t}.
\end{eqnarray}

\bigskip

\noindent \textbf{Utility and Verification.} \added{Individuals in state $I$ expect a (present value) lifetime utility of $1$ if their opinion coincides with the true state of the world and $0$ otherwise. We set the utility of being in state $S$ to $0$.}

\added{As mentioned above, when individuals first hear a message, say $m$, they choose how much effort to exert in verification depending on their belief the message they received is correct. We explain how beliefs are revised below. Exerting effort $\alpha\in [0,\infty)$ implies that verification is successful with probability $x(\alpha)$, in which case the individual knows that $\Phi=0$ for sure. With the complement probability, the result of the verification process is inconclusive.}

An individual of type $b$ has a prior that $b=\Phi$ of $y>.5$. \added{In subsection \ref{endogenous-v}, we derive the threshold of $y$ above which it is optimal for an individual to hold the belief that $b=\Phi$ if her verification of a message is unsuccessful.}

\added{Each agent, being infinitesimal, takes as given the verification levels and information prevalence in the population. Hence,} the utility of individual $i$ who hears message $m$ at time $t$ is
\begin{equation}
 U_{it}=x(\alpha_{it}) +(1-x(\alpha_{it}))Pr_{t}(b=\Phi|m) - c\alpha_{it}, \label{utility}
\end{equation}
where $x(\alpha_{it})$ is the probability verification is successful given a verification effort $\alpha_{it}$, $Pr_{t}(b=\Phi|m)$ is $i$'s expectation of being correct conditional on the message heard and $c$ is the marginal costs of verification. \added{As individuals observe only one message, this expectation is formed updating one's prior using the expected prevalence of messages from the diffusion process described by equations \eqref{rho00}-\eqref{rho11}.}

\added{We denote by $\Delta (x(\alpha))$ the subderivative of $x(\alpha)$, a correspondence that maps for any $\alpha>0$ the values between the right and left derivatives of $x$, and by $g$ the inverse of the subderivative of $x$. With some abuse of notation, we write $\Delta (x)$ instead of $\Delta (x(\alpha))$ when no confusion may arise.}
\begin{assume}\label{H-1}
\added{We assume $x(\cdot): \mathcal{R}^{+}\rightarrow [0,\bar{x}] \subseteq [0,1]$ is strictly increasing and strictly concave on $[0,\bar{x})$, continuous and such that $x(0)=0$.} 
\end{assume}
\added{We denote by $\bar{x}$ and $\bar{d}$ the values such that $ \bar{x} \equiv \displaystyle \lim_{\alpha \rightarrow \infty} x(\alpha)$ and $\Delta (0)=[\bar{d} , \infty)$. Finally,} we denote by $\alpha^b_{it}$ the equilibrium effort individual $i$ of group $b$ exerts in verifying a message in line with their type, i.e., $m=b$, and by $\alpha^{-b}_{it}$ the equilibrium effort to verify $m\neq b$. \added{These efforts lead to the verification rates $\ell_{it}=x(\alpha^b_{it})$ and $h_{it}=x(\alpha^{-b}_{it})$ we employed in equations \eqref{rho00}-\eqref{rho11}.}


\bigskip

\noindent \textbf{Steady State and Equilibrium.} \added{The model is in steady state if equations \eqref{rho00}, \eqref{rho01} and \eqref{rho11} are equal to zero. A steady state of the continuous dynamic system defined by these equations is \emph{locally stable} if it satisfies \emph{Lyapunov stability}.}\footnote{\added{Formally, a steady state is locally stable if, for each neighborhood $S$ of the steady state prevalence of messages, $\rho_{0}^{0}$, $\rho_{0}^{1}$ and $\rho_{1}^{1}$, there exists a neighborhood $W$ such that each
trajectory starting in $W$ remains in $S$, for all $t \geq 0$ and the corresponding trajectory converges to the steady state as $t\rightarrow \infty$.}} \added{A steady state is \emph{positive} if the associated proportion of informed individuals, $\iota$, is strictly positive. We remove the time subscript $t$ to indicate the steady state value of variables.}

\added{The profile of verification efforts $(\alpha^b,\alpha^{-b})$ is an \emph{equilibrium} if it maximizes \eqref{utility} for all individuals taking as given the steady state diffusion rates of messages.}

\bigskip

\noindent \textbf{Discussion and interpretation.} \added{Before presenting the analysis of the model, we discuss its main assumptions. First, we assume that individuals' prior $y$ about the state of the world is sufficiently high that absent verification, they believe their bias to be correct. We focus on this case because, if individuals have lower priors, they believe whichever message they first receive. In this case, rumors either die out, or verification is completely absent. Section \ref{endogenous-v} provides more details.}

\added{A key assumption of our model is that individuals are ``partially Bayesian'': once they first receive a message, they calculate their posterior belief that $b=\Phi$ using Bayes' rule; however, after they have formed their opinion, they do not further update the probability this is correct. One interpretation of this behavior is that after forming an opinion, individuals make a once-for-all decision. While these decisions are not always irreversible, the cost of making a wrong choice are either very substantial and/or realized only after a long delay (such as using a condom, or vaccination decisions). Another interpretation is behavioral: insofar that holding an opinion shapes one's identity, the perceived or psychological cost involved in changing identity may too high with respect to the benefits at stake.}\footnote{\added{Such behavior would be consistent with people filtering out negative information that contradicts their point of view or systems of beliefs in order to maintain a congruent view of the world, and hence their well-being \citep{taylor1988illusion}. This results into information avoidance, inattention, or a biased interpretation of information \citep{golman2017information}. Additionally, identity is rather fixed, as stressed in the literature on inter-group conflict in social psychology \citep{stephan2017} or identity in economics \citep{akerlof2000}.}} \added{In other words, in order to reduce the cognitive dissonance between one's belief and new information acquired in subsequent communication, individuals interpret the latter as supportive of their own opinion. This interpretation is supported by evidence of confirmation bias in online social platforms \citep{zollo2017}.}

\added{Contrarily to the works in the social learning literature using variations of the \cite{deGroot} model, individuals do not exchange opinions with all their neighbors at every time period. Rather, $\nu$ is such that they never receive more than one message per period, and they decide then whether to verify and what to believe. Furthermore, as in \cite{banerjee1993economics}, the message space is coarse, in the sense that a message contains only someone's action or opinion, i.e., $m=0$ or $m=1$, and not the probability they attach to their opinion being correct. This captures the idea that only actions are observable (and not beliefs) or that people transmit only imprecise information regarding their beliefs. An alternative way to model information diffusion would be for individuals to communicate their opinion about the true state of the world, i.e., $Pr(\Phi=0)$. However, in this case verification would automatically become certifiable, as (only) successful verification leads to a posterior belief of $Pr(\Phi=0)=1$. Lastly, individuals do not observe their neighbors' type. If individuals observed their neighbors' opinions or types, or information were certifiable, the rumor would always die out \citep{winner}.}

\added{Finally, we make a number of simplifying assumptions to ease the exposition which are without loss of generality. In particular, our results are qualitatively unaffected by assuming a non-degenerate degree distribution $P(k)$. Our assumption that individuals in state $S$ receive a payoff of $0$ does not affect marginal considerations as transitioning into and out of this state is not a choice. Lastly, our results also hold in the limit of the death rate $\delta\rightarrow 0$. The only adjustment required is to assume that payoffs accrue at a finite time $T$. Thus, we are able to capture the evolution of rumors which diffuse over whole lifespans (such as the HIV-AIDS or vaccination-autism links) as well as more short-lived rumors that diffuse in a constant population.}


\section{Main Analysis} \label{main}

\added{In this section, we solve for the equilibrium of our model. To do so, we proceed in two steps. First, we derive the steady state prevalence of truth and rumor in the population for given verification efforts. This reveals that both have positive prevalence in steady state. Second, we solve for equilibrium verification rates and show that: \textit{(i)}, they depend only on whether the message received is in line or not with one's bias and, \textit{(ii)}, if one's prior $y$ is sufficiently high, it is optimal to hold an opinion in line with the prior if verification is not successful. Finally, we study how the truth to rumor ratio changes with homophily, and how this depends on the verification technology.}

\subsection{Steady State with Exogenous Verification}

\added{We focus now on the model with given verification efforts to understand the properties of the steady state of our model.}

We introduce the \textit{effective diffusion rate}, $\lambda=\nu/\delta$, which summarizes the effect of $\nu$ and $\delta$. Denote by $\iota^b$ the proportion of type $b\in\{0,1\}$ individuals in state $I$ (irrespective of their opinions) with $\iota=(\iota^0+\iota^1)/2$ as overall information prevalence. With this notation in place, we can state the following.
\begin{prop}\label{uniqueness}
\added{Assume verification rates are given. If $\lambda k>1$, there exists a unique positive steady state of the economy; this is locally stable, with $\iota^0=\iota^1=\iota=1-1/(\lambda k)$. If $\lambda k\leq 1$, only a steady state in which the prevalence of both the truth and the rumor are zero is locally stable.}
\end{prop}
While there always exists a steady state in which the prevalence of both the truth and the rumor are zero---if no individual ever transmits an information, nobody can ever become informed,---this is stable only if $\lambda k\leq 1$. Otherwise, there is a unique positive steady state, and it is stable.

Proposition \ref{uniqueness} establishes information prevalence within each group. However, this does not tell us the diffusion of the truth and the rumor. This is the object of the following Proposition.
\begin{prop}\label{exogenous}
\added{In the unique \added{positive and locally} stable steady state, for exogenous verification rates $\ell$ and $h$:\\
\textit{i)} the information prevalence of both messages, $\rho_0$ and $\rho_1$, is increasing in the effective diffusion rate, $\lambda$, and in the number of meetings, $k$;\\
\textit{ii)} the truth to rumor ratio, $\rho_0/\rho_1$, is greater than $1$, increasing in both verification rates, $\ell$ and $h$, and independent of the effective diffusion rate, $\lambda$, and the number of meetings, $k$;\\
\textit{iii)} the truth to rumor ratio, $\rho_0/\rho_1$, is decreasing in homophily, $\beta$, if and only if individuals verify more a message against their bias, i.e., $h>\ell$.}
\end{prop}
Proposition \ref{exogenous} results from the steady state prevalence of truth and rumor:
\begin{eqnarray}
\rho_0 &=& \frac{1}{2} \cdot \frac{1+h- 2 \beta (h - \ell)}{1- \beta (h - \ell)}\iota, \label{SSrho0}\\
\rho_1 &=& \frac{1}{2} \cdot \frac{1-h}{1- \beta (h - \ell)}\iota. \label{SSrho1}
\end{eqnarray}
These equations show that the steady states of opinions inherit uniqueness and stability from $\iota$. Since for $\lambda k \leq 1$, neither opinion is endemic, we focus in the remainder of the paper on $\lambda k> 1$. Equation \eqref{SSrho1} highlights that both opinions survive unless $h=1$. Thus, rumors may survive in the long run even if verification is possible.

Equations \eqref{SSrho0} and \eqref{SSrho1} show that with zero verification effort, $\rho_0=\rho_1=\iota/2$, and the truth prevalence increases in any form of verification, while rumor prevalence decreases. Hence, for any positive amount of verification, the truth exhibits a larger prevalence than the rumor.

Proposition \ref{exogenous} delivers some insights about potential relationships between online social networks and the diffusion of rumors. One factor through which online social networks allegedly stimulate the diffusion of rumors is the ease with which messages can be communicated, and the number of people receiving them. Thus, one generally expects them to have increased $k$, $\nu$, or both. Proposition \ref{exogenous} shows that our model's predictions are in line with this view. However, it stresses that the truth and the rumor equally benefit from an increase in the ease of communication due to an increase in the number of meetings $k$ or in the effective diffusion rate $\lambda$. 

This result has several implications. First, while empirical studies on the impact of online communication often focus on the diffusion of rumors alone (e.g., \citealp{zollo2017}), Proposition \ref{exogenous} stresses that a comparison with the diffusion of truthful messages would be of a greater interest. \added{In line with this insight, the truth to rumor ratio will be the main object of interest in the remainder of the paper.}

Second, if rumors have indeed become more prevalent in relative terms, this cannot be explained by online social networks increasing communication rates \textit{per se}. We discuss in Section \ref{policies} how ease of communication might have indirect effects on relative rumor prevalence.

\added{Likewise, high degrees of homophily are commonly associated with an increased diffusion of rumors as people are likely to hear only messages in line with their bias (``echo chambers'').} In fact, in our model the impact of homophily on the diffusion of truth and rumor depends entirely on the verification rates of messages. \added{Fixing these rates, homophily indeed benefits the rumor and harms the truth if individuals are more likely to verify messages against their bias than those aligned with it. While such behavior appears intuitive, it motivates us to study endogenous verification next.}

\subsection{Endogenous Verification} \label{endogenous-v}

\added{Given the utility function \eqref{utility}, each individual $i$ chooses a verification effort when they first hear a message such that}
\begin{equation} \label{FOC-interior}
 \added{g \left( \frac{c}{1-Pr_{t}(b=\Phi|m)} \right) = x(\alpha_{it}),}
\end{equation}
\added{where $g$ is the inverse of the subderivative of $x$.}

\added{Individuals hence need to calculate the probability that $b=\Phi$ conditional on having received $m$. We assume that they perform this calculation using Bayes' rule, given that they are aware of the transmission process and that their prior of being of type $b=\Phi$ is $y$. This leads to the following:}
\begin{eqnarray}
 Pr_t(b=\Phi|m=b) & = &
 \added{\frac{y(\beta\rho^0_{0,t}+(1-\beta)\rho^1_{0,t})}{y(\beta\rho^0_{0,t}+(1-\beta)\rho^1_{0,t})+(1-y) \beta\rho^1_{1,t}}}, \label{equ-probmb}
\\
 Pr_t(b=\Phi|m\neq b) & = &
 \added{\frac{y(1-\beta)\rho^1_{1,t}}{y(1-\beta)\rho^1_{1,t}+(1-y)((1-\beta)\rho^0_{0,t}+\beta\rho^1_{0,t})}}.\label{equ-probmnb}
\end{eqnarray}

\added{From these probabilities and \eqref{FOC-interior}, equilibrium verification effort depends on whether the message received is in line or not with one's bias, and not on the prior \textit{per se}. Therefore, in equilibrium there are two verification rates. This result follows from two properties of our model. First, all individuals believe their bias is correct with the same probability, $y$. Second, individuals derive prevalence of messages from the diffusion process described by equations \eqref{rho00}, \eqref{rho01} and \eqref{rho11}, as they do not observe all their neighbors' opinions when they set their verification effort. Substituting \eqref{equ-probmb}, \eqref{equ-probmnb}, \eqref{SSrho0} and \eqref{SSrho1} in \eqref{FOC-interior}, equilibrium verification efforts are thus described by}
\begin{align}
 \ell = x (\alpha^b) & = \added{g \left( \frac{c (y (\beta + (1-\beta) h - \beta (h-\ell)) + (1 - y) \beta (1 - h))}{(1-y)\beta(1-h)} \right)},\label{equb} \\
 h = x ( \alpha^{-b} ) & = \added{ g \left( \frac{c(y(1-\beta)(1-h)+(1-y)(1-\beta+\beta \ell))}{(1-y)(1-\beta +\beta \ell) } \right) }. \label{equ-b}
\end{align}
We now prove that an equilibrium with endogenous verification exists.
\begin{prop}
\label{Existence}
\added{Under Assumption \ref{H-1}, there exists a threshold $\overline{y}$ such that, if $y\geq\overline{y}$, an equilibrium of the model with laws of motion \eqref{rho00}, \eqref{rho01}, \eqref{rho11} and endogenous verification exists.}
\end{prop}
\added{Holding an opinion in line with one's bias when a message is not verified, as we assumed so far, is optimal if $Pr(b=\Phi|m\neq b)\geq .5$. By \eqref{equ-probmnb}, in steady-state this requirement translates into}
\begin{equation}\label{cond-y}
 \added{\frac{y}{1-y}\geq \frac{1-\beta+\beta\ell}{(1-\beta)(1-h)}}.
\end{equation}
\added{We show that there exists a threshold $\overline{y}$ such that \eqref{cond-y} is satisfied if $y\geq\overline{y}$. Intuitively, if one's prior is sufficiently strong, lacking verification, informed individuals hold an opinion in line with their bias.}

\added{If condition \eqref{cond-y} is not met, absent verification, players believe the first message they hear. In that case, there cannot be positive verification rates in steady state. If the prevalence of the rumor is such that it is worthwhile to verify, people do so, thereby reducing this prevalence, until verification is no longer profitable. At that point, the truth to rumor ratio stays constant, which might entail the prevalence of the rumor to be infinitesimally small.}

\added{The following propositions characterizes equilibrium verification rates.}
\begin{prop}\label{verif}
\added{If Assumption \ref{H-1} holds and $y\geq\overline{y}$, in any equilibrium, individuals exert higher effort verifying messages against their bias than those in line with it ($\ell\leq h$). Equilibrium verification is independent of the number of meetings, $k$, and the effective diffusion rate, $\lambda$. Furthermore, there exist values on verification costs $\underline{c}$ and $\overline{c}$ such that} any equilibrium takes one of the following forms:\\
\textit{i)} If $c\geq \added{\overline{c}}$, there is no verification and the truth to rumor ratio is 1.
\\
\textit{ii)} If $c\in[\underline{c}, \added{\overline{c}})$, individuals verify only messages against their bias.
\\
\textit{iii)} If $c<\underline{c}$, both messages are verified.
\\
\added{While both $\overline{c}$ and $\underline{c}$ are decreasing in $y$, $\overline{c}$ is independent of homophily, $\beta$, and $\underline{c}$ increasing in it.} \added{Finally, the corresponding steady state is locally stable} if and only if, either it coincides with a zero steady state and $\lambda k\leq1$, or it is positive and $\lambda k > 1$.
\end{prop}
\added{The intuition behind these results is the following. First, a message in line with one's bias is verified less than one against it, because receiving the latter implies a lower probability of one's bias to be correct after Bayesian updating.}

When verification is very costly, no message is verified; as a result, there is an equal mass of individuals holding each of the two opinions. As verification costs decrease, individuals first verify the message against their bias, \added{as $\ell\leq h$. For even lower verification costs, individuals verify both messages. Verification implies the truth has a higher prevalence than the rumor.}

The threshold of verification cost below which both messages are verified is increasing in homophily. Indeed, when individuals are more likely to meet people with the same bias, they attach a lower informational content to messages in line with their bias, thereby triggering increased verification.

\added{Finally, Proposition \ref{verif} provides conditions for the stability of the steady state in the endogenous equilibrium that come directly from Proposition \ref{uniqueness}, as these conditions are independent of those determining verification.}

\added{We present a simple and intuitive example of a verification function that admits explicit solutions. All computations are in Appendix \ref{app:examples}.}

\begin{example}[Exponential verification function with a cap]
\label{ex:exp_with_cap}
\normalfont \added{Let the verification function be:}
\begin{eqnarray}\label{function:exp}
x (\alpha) = \left\{ 
\begin{array}{lll}
1-e^{-\alpha} & \mbox{ if } \alpha < -log[1-\bar{x}] , \\
\bar{x} & \mbox{ if } \alpha \geq -log[1-\bar{x}]. \\
\end{array}
\right.
\end{eqnarray}
\added{This function results from the following verification process. Consider an individual who searches for information, which consists of $n$ realizations leading to an answer with probability $p_n$ per realization, up to the point where all information available, $\bar{x}$, is collected. Before reaching this cap, this search process gives at least one answer with probability $1-(1- p_n/n)^{n}$. If $n p_n=\alpha$ as the number of realizations $n$ goes to infinity {and the success of each realization goes to zero}, this probability converges to $1-e^{-\alpha}$ if this is lower than $\bar{x}$, and $\bar{x}$ otherwise, leading to \eqref{function:exp}.}\footnote{Realizations might have multiple answers. In the context of this model, this is equivalent to a lower cost of effort in verification.}

\added{As $\bar{x}\rightarrow 1$, \eqref{function:exp} converges to $1-e^{-\alpha}$. In this case, Figure \ref{fig:exp} shows which messages are verified, depending on parameters $y$ and $c$, for $\beta=0.6$. The figure shows that, if verification costs are too high, no message is verified (the light-blue region). When the costs are sufficiently low, we have two scenarios. If the prior $y$ is relatively low, both messages are verified (the purple region), but if it is higher, only the message against one's bias is verified (the blue region). Additionally, the initial prior $y$ has to be above the threshold $\bar{y}$ derived in Proposition \ref{Existence}: in the white region, this condition is not satisfied.}

\begin{figure}[ht!]
 \centering
 \includegraphics[width=12cm]{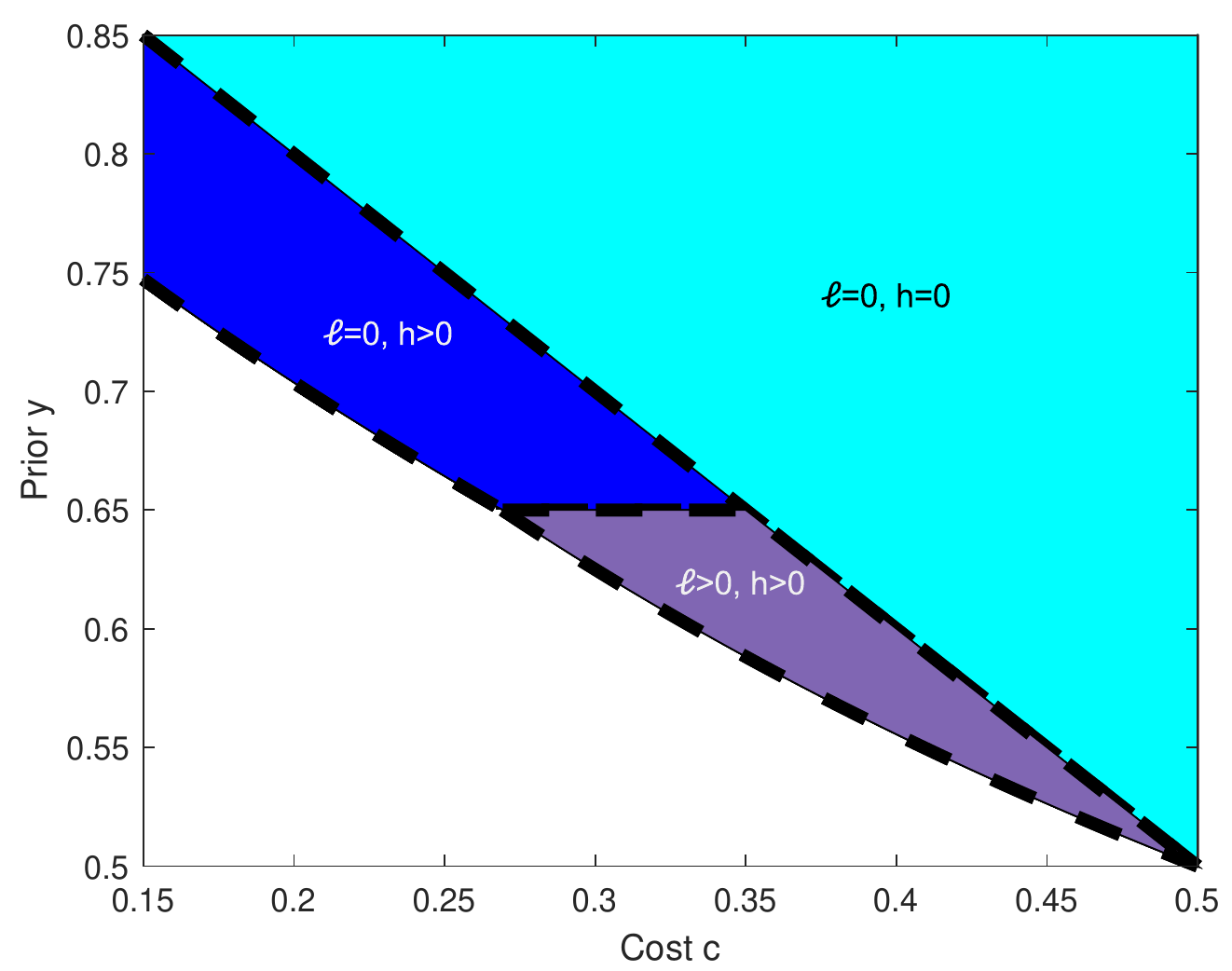}
 \caption{Regions of the parameters $c$ (on the $x$-axis) and $y$ (on the $y$-axis) where different kind of equilibria exist with the exponential verification function, $\bar{x}\rightarrow 1$ and $\beta=0.65$.}
 \label{fig:exp}
\end{figure}

\added{If instead $\bar{x}\in (0,1)$, there are regions in the $(c,y)$ space where verification rates are at $\bar{x}$. We will discuss this case in greater detail in the next subsection, as the value of $\bar{x}$ is important to understand the effect of homophily on the truth to rumor ratio. }

\end{example}


\subsection{Truth to rumor ratio}

The truth to rumor ratio is the fraction of the prevalence of both opinions among informed individuals, i.e., $\rho_0/\rho_1$. It derives from equations \eqref{SSrho0} and \eqref{SSrho1} as follows
\begin{equation}
 \frac{\rho_0}{\rho_1}=\frac{1+h}{1-h}- 2 \beta \frac{h - \ell}{1-h}.\label{truthtorumor}
\end{equation}
\added{Equation \eqref{truthtorumor} highlights that homophily has two effects on the truth to rumor ratio. First, there's a direct effect: as homophily increases, individuals are more exposed to messages in line with their bias, which, as Proposition \ref{verif} shows, are verified less. Hence, the truth to rumor ratio decreases.}

\added{However, endogenous debunking implies an indirect effect through equilibrium behavior: as homophily increases, the informativeness of messages in line with one's bias decreases, precisely because agents are more exposed to messages that are verified less. The opposite for messages against one's bias. In both cases, the probability they attach to their bias being correct decreases. As a result, individuals verify more, which increases the truth to rumor ratio.}

\added{Which of the two effects dominates is \textit{a priori} not clear. Proposition \ref{ttrr} however derives two insights in this respect. Denote $L\geq 0$ as $g(L)=\ell$; hence, $L$ represents the optimal marginal increase in the verification success of messages in favor of one's bias, $x(\alpha^b)$.}
\begin{prop}\label{ttrr}
In equilibrium, the truth to rumor ratio, $\rho_0/\rho_1$, is equal to
\[
\label{truthtorumor2}
\added{1+ \frac{2}{y} \left(\frac{L(1-y)}{c} -1 \right) \beta .}
\]
Furthermore, $\rho_0/\rho_1$ is decreasing in homophily, $\beta$, if $c\in[\underline{c},\added{\overline{c}}]$, i.e., when individuals verify only messages against their bias. 
\end{prop}
\added{Proposition \ref{ttrr} first shows that, whenever only messages against one's bias are verified, the indirect effect is absent. Indeed, we show in the Appendix that $h$ does not change with homophily when $l=0$. As a result, for verification costs such that this is an equilibrium, the truth to rumor ratio is decreasing in homophily.}

\added{Second, when both messages are verified, $L$ is a sufficient statistic to study the truth to rumor ratio.
In particular, Proposition \ref{ttrr} allows us to show that the total effect of homophily on the truth to rumor ratio depends on the local concavity of the verification function.}

\added{Indeed, if the verification function $x(\alpha)$ is twice differentiable, we find that the total effect of homophily on the truth to rumor ratio is}
\begin{eqnarray*}
\added{\frac{d}{d \beta} \frac{\rho_0}{\rho_1} =
\frac{2}{y} \left(\frac{L (1-y) }{c} -1 \right)+\frac{2 \beta (1-y)}{c y} \frac{d L}{d \beta}} \label{homophilyontruth}
\end{eqnarray*}
\added{Hence, the effect is positive if}
\begin{eqnarray}
\added{L\geq \frac{c}{1-y}+\beta \left\vert \frac{d L}{d \beta} \right\vert .}\label{total_effect_hom}
\end{eqnarray}
\added{While $L$ is equal to $dx(\alpha)/d\alpha$ evaluated at $\alpha^{b}$, i.e., depends on the steepness of the verification function $x( \alpha )$, $dL/d\beta$ depends on its concavity, as}
\[ \added{\frac{d L}{d \beta} = \frac{\partial \ell}{ \partial \beta} \cdot \left. \frac{\partial^2 x( \alpha )}{ \partial \alpha^2} \right|_{x(\alpha)=\ell}.}
\]
\added{Hence, when homophily increases, if verification increases faster than the decline in its marginal benefits, i.e., $L$ is sufficiently larger than $dL/d\beta$, verification increases sufficiently that the truth to rumor ratio increases as well.}

\added{The role that steepness and concavity of the verification function play highlights why homophily may affect the truth to rumor ratio in non-obvious ways. We further elaborate on these results using the verification function we introduced in Example \ref{ex:exp_with_cap}. All computations are in Appendix \ref{app:examples}.} 

\begin{example}[Effects of homophily in Example \ref{ex:exp_with_cap}]
\label{ex:marginal_homphily}
\normalfont
\added{By Proposition \ref{ttrr}, it is interesting to study the effect of homophily on the truth to rumor ratio only whenever $y\geq\bar{y}$ and $c<\underline{c}$, such that both messages are verified.}

\added{In the case that $0<\ell<h=\bar{x}$, if verification costs are low, so that $\ell$ is large (but below $\bar{x}$), the exponential function is not as concave as it is near the origin, and $\left| d L/d \beta \right|$ is low enough to satisfy inequality \eqref{total_effect_hom}.
As verification costs increase, $\ell$ goes down, $\left| d L/d \beta \right|$ goes up, and inequality \eqref{total_effect_hom} is reversed. As verification costs increase further, also $h$ is less than $\bar{x}$, i.e., $0<\ell<h<\bar{x}$. We discussed this case in Example \ref{ex:exp_with_cap} (purple region in Figure \ref{fig:exp}). In this case, the truth to rumor ratio becomes $(2(1-y)-c)/c$, which is independent of homophily. This is due to a specificity of the exponential example, where, when both verification rates can freely adapt, the direct and the indirect effects of homophily balance each other out. Note that, in all regions verification rates and the truth to rumor ratio are (weakly) decreasing in verification costs.}

\begin{figure}[htb!]
 \centering
 \includegraphics[width=12cm]{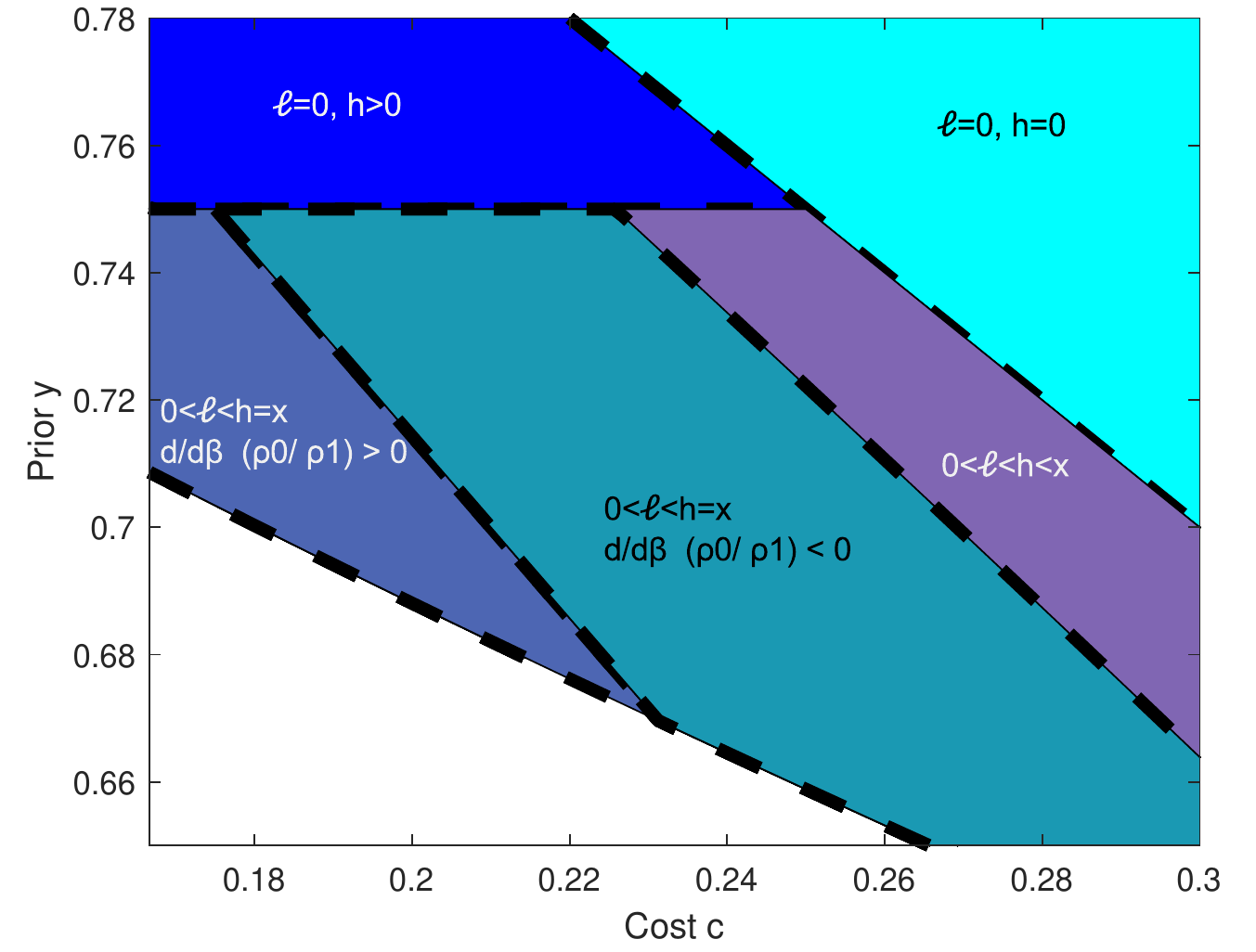}
 \caption{Regions of the parameters $c$ (on the $x$-axis) and $y$ (on the $y$-axis) where different kind of equilibria exist with the exponential verification function, $\beta=0.75$ and a cap set at $\bar{x}=0.3$.}
 \label{fig:exp_with_cap}
\end{figure}

\added{Figure \ref{fig:exp_with_cap} depicts these regions, illustrating how the effect of homophily on the truth to rumor ratio differs depending on the values of the prior $y$ and verification costs, $c$.}
\end{example}

\added{In sum, even in a simple and homogeneous verification function as the one assumed here, the impact of homophily on the truth to rumor ratio depends on exact equilibrium reactions, which are difficult to predict. Despite this, our model delivers several policy implications, which we discuss in the following section.}

\section{Discussion and Policy Implications} \label{policies}

Much of the ongoing discussion on how to fight the spread of rumor rests on the role of online social networks. In the following, we address the most commonly cited arguments as to how these platforms relate to the diffusion of rumors.

\subsection{Time-limited Injections of Messages}

\added{One obvious policy in the fight against rumors is to increase the volume of truthful messages in the network for a limited time, e.g., through an information campaign. In our model, whenever the verification function delivers a unique stable equilibrium, such a policy does not affect the truth to rumor ratio in the long run. Thus, rumors cannot simply be debunked by increasing the prevalence of the truthful message in the network for a limited time.}

\subsection{Online Social Networks Display Higher Homophily}

\added{In principle,} online social networks offer individuals the chance to self-select in more homogeneous groups along several dimensions than offline communication, as meeting opportunities are less constrained online. \added{The extent to which this happens is still debated. On the one hand, there is evidence that patterns of online news consumption are no more segregated or polarized than offline ones \citep{Gentzkow,boxell2018}. On the other hand, online communication patterns display levels of segregation that are above the most segregated offline communication networks found by \cite{Gentzkow}, e.g., in \cite{Halberstam} for Twitter and \cite{zollo2017} for Facebook. As a result, the widespread use of online social networks might have implied an increase in homophily and resulted in echo chambers.} It is usually assumed that this is one of the main culprits for an increase in the diffusion of rumors in the past decade.

\added{Granting that online social networks have increased homophily,} our preceding analysis makes the mechanism behind this argument explicit, stressing under which conditions it is valid. \added{As meetings are more homophilous, on the one hand, individuals hear fewer discordant messages, that are verified less. On the other hand, messages confirming one's prior become less informative, so that these messages are verified more. Rumors thrive only when this first effect dominates.}

\subsection{Online Social Networks Facilitate Communication}

\added{By facilitating communication, online social networks have arguably increased the transmission rate of information. Through the lens of our model, this is captured by an increase in either the number of meetings, $k$, or in the diffusion rate, $\lambda$. This} increases the measure of informed individuals, but it does not affect verification rates (Proposition \ref{verif}).\footnote{In a more general model, the same holds for changes in the degree distribution $P(k)$.} The truth to rumor ratio is then unaffected by changes in these parameters. Therefore, while ease of communication leads to an absolute increase in rumor prevalence, this in itself does not imply a \textit{relative} increase.

It is possible, however, that the costs of verification, $c$, instead depend positively on the amount of online communication. Such \textit{congestion} or \textit{information overload} effects occur naturally if we consider verification costs as time costs. In that case, through increasing communication, online social networks may increase verification costs and \added{thus affect verification rates.}

To fix ideas, consider individuals to have a given time endowment $\mathcal{H}$ in each period, which they can spend on communication ($\nu k$), verification ($\alpha$), and on independent leisure activities $z$. Assume that agents obtain utility $u(z)$ from their leisure activities, where $u(\cdot)$ is a continuous, increasing and concave function. Its concavity captures the decreasing marginal utility of private leisure. Agent $i$'s utility at time $t$ can then be expressed as

\begin{equation}
 U_{it} = x(\alpha_{it}) + \left(1-x(\alpha_{it})\right)Pr_i(b=\Phi|m) + u(\mathcal{H}-\nu k - \alpha_{it}).
\end{equation}
Optimal verification effort is chosen again according to equation \eqref{FOC-interior}, defining $c=du/d\alpha_i$. In this case, any increase in $\nu k$ increases verification costs.

We conclude that if increased communication creates congestion effects in verification, online social networks may indeed have \added{affected verification rates and therefore the truth to rumor ratio.}

\subsection{Fact Checking to Fight Rumors}

The ease with which rumors can be debunked by agents in the network has been an important aspect in discussions on policy interventions. Many established news outlets, such as \textit{The New York Times} or \textit{Le Monde}, publish guides on how to recognize false information, and have introduced newsrooms where fake news is identified and debunked \citep{Roose}. On various online social networks, for example \textit{Facebook}, disputed information may be ``flagged" as such \citep{facebook} by independent fact checkers.

\added{In our model, incentives that induce higher verification rates unambiguously increases the truth to rumor ratio. Hence, fact checking should be properly designed.}

We believe this is not always the case. For example, Facebook experimented with assigning ``flags" to posts simply stating that they had been disputed. This practice has been abandoned in $2017$, as it proved less effective to stop the spread of rumors than expected, with some evidence that it might even have promoted the spread of such posts. Instead, an alternative was suggested whereby, next to a disputed post, a user would see various links to the articles disputing it \citep{Silverman}. In the context of our model, the earlier policy would have amounted to injecting truthful messages into the network, and as such would indeed have been expected to be unsuccessful in the long run. The updated policy instead, by providing links to the sources of the dispute, can credibly claim to lower verification costs. 

Overall, our model corroborates that one of the most intuitive policies to deter rumors---incentivizing individuals to verify information---is also the one most likely to succeed as long as it is carefully designed.

\section{Partisans}
\label{sec:partisan}

It is possible that the diffusion of rumors in social media is primarily driven by individuals that hold a certain opinion independently of whether they are confronted with factual evidence disproving it \citep{zollo2017}. We now study the implication of the existence of such partisan individuals on the diffusion of messages.

We introduce partisans as people who never verify, always hold an opinion in line with their prior and transmit messages accordingly.\footnote{Alternatively, for all $\beta<1$, we can think of partisans as having a prior of $1$ of being biased towards the true state of the world. If $\beta=1$, individuals can never meet someone with the opposing bias. Hence, hearing the opposite message is conclusive proof that one's bias is wrong, contradicting the certain prior.}

We assume that a fraction $\gamma$ of individuals of each type are partisans. This does not impact the evolution of information in group $0$, as both partisans and non-partisans are always of the opinion that $\Phi=0$. However, there are fewer individuals of type $1$ who may verify any message they hear. We then derive the following Proposition.
\begin{prop} \label{partisan}
Assume a fraction $\gamma$ of the population consists of partisans. The prevalence of the truth and the rumor among the remaining $1-\gamma$ individuals are unaffected by $\gamma$ and remain as in the baseline model.
\end{prop}
This result depends on the endogeneity of verification. Indeed, the existence of partisans implies that non-partisans place a higher probability of being told the rumor. This therefore leads to a higher degree of verification among them, which offsets the negative impact that partisans have on the relative prevalence of truth and rumor. Formally, in the proof of Proposition \ref{partisan} we show that the prevalence of both opinions can be re-written as in the baseline model with verification rates $\hat{x}(\alpha)=(1-\gamma)x(\alpha)$ instead of the original $x(\alpha)$'s. This leads to the equilibrium $\hat{x}(\alpha)$ being unaffected by partisans.


\section{Conclusions}\label{conclusions}

In this paper, we model how a correct message and a rumor diffuse in a population of individuals who seek the truth. Individuals verify the messages they hear based on the probability that what they hear is correct. They are biased in a way that, if they do not verify, they hold the opinion that adheres to their view of the world.

We find that the rumor survives for any positive cost of verification. New communication technologies increase its absolute prevalence, however, its prevalence relative to the truth depends exclusively on verification and the degree of homophily in meetings. \added{We show that the impact of homophily is nuanced. On the one hand, an echo chamber effect emerges: individuals are more exposed to messages in line with their bias, which are verified less. On the other, higher homophily means messages reinforcing ones prior become less informative, and the opposite for messages going against one's prior. This mechanism then leads to higher verification rates. Overall, the relative virality of rumors increases only when the first effect dominates the latter.}

We employ our results to discuss policies to debunk rumors. While injections of truthful messages are ineffective in debunking the rumor in the long run, successful policy interventions revolve around incentivizing individuals to verify. In the light of our model, lowering the degree of homophily may fail to achieve this result. In sum, new communication technologies played a role in making rumors more viral if they decreased verification---due to, for example, congestion effects.

In our model, there only are two opposing opinions and messages. This excludes the possibility that malevolent agents may aim to decrease the spread of the truth by creating new opposing messages over time. \added{We think the analysis of this phenomenon is a promising avenue for future research.}

\FloatBarrier

\bibliographystyle{aea.bst}
\bibliography{bibdebunk}

@article{bizzarri2020segregating,
  title={Epidemic dynamics with homophily, vaccination choices, and pseudoscience attitudes},
  author={Bizzarri, Matteo and Panebianco, Fabrizio and Pin, Paolo},
  journal={arXiv preprint arXiv:2007.08523},
  year={2021}
}

@article{bolletta2019polarization,
  title={Polarization when people choose their peers},
  author={Bolletta, Ugo and Pin, Paolo},
  journal={Available at SSRN 3245800},
  year={2021}
}

@article{samantray2019credibility,
  title={Credibility of climate change denial in social media},
  author={Samantray, Abhishek and Pin, Paolo},
  journal={Palgrave Communications},
  volume={5},
  number={1},
  pages={1--8},
  year={2019},
  publisher={Nature Publishing Group}
}

@article{Allcott,
  title={Social media and fake news in the 2016 election},
  author={Allcott, Hunt and Gentzkow, Matthew},
  journal={Journal of Economic Perspectives},
  volume={31},
  number={2},
  pages={211--36},
  year={2017}
}

@techreport{clz19,
	Author = {Campbell, Arthur and Leister, C. Matthew and Zenou, Yves},
	Institution = {Mimeo},
	Title = {Social media and polarization},
	Year = {2019}}

@techreport{akbarpour2018,
  Author={Akbarpour, Mohammad and Malladi, Suraj and Saberi, Amin},
  Title={Just a few seeds more: value of network information for diffusion},
  Institution={Mimeo},
  Year={2018}
}

@techreport{kr20,
	Author = {Kranton, Rachel and McAdams, David},
	Institution = {Mimeo},
	Title = {Social networks and the market for news},
	Year = {2020}}

@article{toxvaerd2019,
  title={Rational disinhibition and externalities in prevention},
  author={Toxvaerd, Flavio},
  journal={International Economic Review},
  volume={60},
  number={4},
  pages={1737--1755},
  year={2019},
  publisher={Wiley Online Library}
}

@article{toxvaerd2014,
  title={The economics of vaccination},
  author={Chen, Frederick and Toxvaerd, Flavio},
  journal={Journal of Theoretical Biology},
  volume={363},
  pages={105--117},
  year={2014},
  publisher={Elsevier}
}

@article{kremer1996,
  title={Integrating behavioral choice into epidemiological models of AIDS},
  author={Kremer, Michael},
  journal={The Quarterly Journal of Economics},
  volume={111},
  number={2},
  pages={549--573},
  year={1996},
  publisher={MIT Press}
}

@article{KM,
	Author = {Kinateder, Markus and Merlino, Luca P.},
	Journal = {American Economic Journal: Microeconomics},
	Number = {3},
	Pages = {187--212},
	Title = {Public goods in endogenous networks},
	Volume = {9},
	Year = {2017}}

@article{Bloch-Rumors,
	Author = {Bloch, Francis and Demange, Gabrielle and Kranton, Rachel},
	Journal = {International Economic Review},
	Number = {2},
	Pages = {421--48},
	Publisher = {Wiley Online Library},
	Title = {Rumors and social networks},
	Volume = {59},
	Year = {2018}}

@article{WarRoom,
	Author = {Sheera Frenkel and Mike Isaac},
	Journal = {The New York Times},
	Month = {Sept. 19, 2018},
	Title = {Inside Facebook's election `war room'},
	Url = {http://www.nytimes.com/2018/09/19/technology/facebook-election-war-room.html},
	Year = {2018}}

@article{Silverman,
	Author = {Silverman, Craig},
	Journal = {BuzzFeed.com},
	Lastchecked = {February, 2019},
	Title = {This analysis shows how fake election news stories outperformed real news on Facebook},
	Url = {https://www.buzzfeednews.com/article/craigsilverman/viral-fake-election-news-outperformed-real-news-on-facebook},
	Volume = {Nov., 16},
	Year = {2016}}

@article{Needle,
	Author = {Shirley Cramer},
	Journal = {Retrieved from https://www.rsph.org.uk/},
	Publisher = {Royal Society for Public Health},
	Title = {Moving the needle: Promoting vaccination uptake across the life course},
	Url ={https://www.rsph.org.uk/uploads/assets/uploaded/9bdd64f9-6b9e-4d93-86561ce4598bc49e.pdf},
	Year = {2018}}

@article{Marr,
	Author = {Marr, Bernard},
	Journal = {Forbes.com},
	Title = {Coronavirus fake news: How Facebook, Twitter, and Instagram are tackling the problem},
	Volume = {March, 27},
	Year = {2020}}

@article{facebook,
	Author = {Maidenberg, Micah},
	Lastchecked = {December 5, 2018},
	Journal = {Retrieved from  https://www.wsj.com/articles/facebook-to-start-fact-checking-photos-videos-1536867288 on December 5, 2018},
	Title = {Facebook to start fact-checking photos, videos},
	Urldate = {September,13, 2018},
	Year = {2018}}

@article{Goyal-Vigier,
	Author = {Goyal, Sanjeev and Vigier, Adrien},
	Journal = {Journal of Public Economics},
	Pages = {64--69},
	Publisher = {Elsevier},
	Title = {Interaction, protection and epidemics},
	Volume = {125},
	Year = {2015}}

@article{mueller2016,
  title={Social learning with costly search},
  author={Mueller-Frank, Manuel and Pai, Mallesh M},
  journal={American Economic Journal: Microeconomics},
  volume={8},
  number={1},
  pages={83--109},
  year={2016}
}

@article{ali2018herding,
  title={Herding with costly information},
  author={Ali, S Nageeb},
  journal={Journal of Economic Theory},
  volume={175},
  pages={713--729},
  year={2018},
  publisher={Elsevier}
}

@article{Bogart,
	Author = {Bogart, Laura M and Thorburn, Sheryl},
	Journal = {JAIDS Journal of Acquired Immune Deficiency Syndromes},
	Number = {2},
	Pages = {213--218},
	Publisher = {LWW},
	Title = {Are HIV/AIDS conspiracy beliefs a barrier to HIV prevention among African Americans?},
	Volume = {38},
	Year = {2005}}

@article{deGroot,
	Author = {DeGroot, Morris H},
	Journal = {Journal of the American Statistical Association},
	Number = {345},
	Pages = {118--121},
	Publisher = {Taylor \& Francis},
	Title = {Reaching a consensus},
	Volume = {69},
	Year = {1974}}

@article{Galeotti-Christian,
	Author = {Galeotti, Andrea and Ghiglino, Christian and Squintani, Francesco},
	Journal = {Journal of Economic Theory},
	Number = {5},
	Pages = {1751--1769},
	Publisher = {Academic Press},
	Title = {Strategic information transmission networks},
	Volume = {148},
	Year = {2013}}

@article{Galeotti-Rogers-AEJ,
	Author = {Galeotti, Andrea and Rogers, Brian W},
	Journal = {American Economic Journal: Microeconomics},
	Number = {2},
	Pages = {1--32},
	Publisher = {American Economic Association},
	Title = {Strategic immunization and group structure},
	Volume = {5},
	Year = {2013}}

@article{Gentzkow,
	author = {Gentzkow, Matthew and Shapiro, Jesse M.},
	doi = {10.1093/qje/qjr044},
	issn = {0033-5533},
	journal = {The Quarterly Journal of Economics},
	month = {11},
	number = {4},
	pages = {1799-1839},
	title = {Ideological segregation online and offline},
	volume = {126},
	year = {2011}
	}

@article{Golub-Jackson,
	Author = {Golub, Benjamin and Jackson, Matthew O},
	Journal = {American Economic Journal: Microeconomics},
	Number = {1},
	Pages = {112--149},
	Publisher = {American Economic Association},
	Title = {Naive learning in social networks and the wisdom of crowds},
	Volume = {2},
	Year = {2010}}

@article{Hagenbach-Koessler,
	Author = {Hagenbach, Jeanne and Koessler, Fr{\'e}d{\'e}ric},
	Journal = {The Review of Economic Studies},
	Number = {3},
	Pages = {1072--1099},
	Publisher = {Oxford University Press},
	Title = {Strategic communication networks},
	Volume = {77},
	Year = {2010}}

@article{Halberstam,
  title={Homophily, group size, and the diffusion of political information in social networks: Evidence from Twitter},
  author={Halberstam, Yosh and Knight, Brian},
  journal={Journal of Public Economics},
  volume={143},
  pages={73--88},
  year={2016},
  publisher={Elsevier}
}

@article{sadler2020,
  title={Diffusion games},
  author={Sadler, Evan},
  journal={American Economic Review},
  volume={110},
  number={1},
  pages={225--70},
  year={2020}
}

@inproceedings{winner,
	Author = {Prakash, B Aditya and Beutel, Alex and Rosenfeld, Roni and Faloutsos, Christos},
	Booktitle = {Proceedings of the 21st international conference on World Wide Web},
	Organization = {ACM},
	Pages = {1037--1046},
	Title = {Winner takes all: competing viruses or ideas on fair-play networks},
	Year = {2012}}

@article{Roose,
	Author = {Kevin Roose},
	Journal = {The New York Times},
	Title = {We asked for examples of election misinformation. You delivered.},
	Url = {https://www.nytimes.com/2018/11/04/us/politics/election-misinformation-facebook.html},
	Volume = {November, 4},
	Year = {2018}}

@article{Tabasso1,
	Author = {Tabasso, Nicole},
	Journal = {Games and Economic Behavior},
	Pages = {219--40},
	Title = {Diffusion of multiple information: On information resilience and the power of segregation},
	Volume = {118},
	Year = {2019}}

@article{zollo2017,
  title={Debunking in a world of tribes},
  author={Zollo, Fabiana and Bessi, Alessandro and Del Vicario, Michela and Scala, Antonio and Caldarelli, Guido and Shekhtman, Louis and Havlin, Shlomo and Quattrociocchi, Walter},
  journal={PloS one},
  volume={12},
  number={7},
  year={2017},
  pages={e0181821},
  publisher={Public Library of Science}
}

@article{akerlof2000,
  title={Economics and identity},
  author={Akerlof, George A and Kranton, Rachel E},
  journal={Quarterly Journal of Economics},
  volume={115},
  number={3},
  pages={715--753},
  year={2000},
  publisher={MIT Press},
  doi= {10.1162/003355300554881}
}

@incollection{stephan2017,
	author = {Stephan, Walter G and Stephan, Cookie White},
	booktitle = {The International Encyclopedia of Intercultural Communication},
	editor = {Y.Y. Kim},
	pages = {1--12},
	publisher = {Wiley Online Library},
	title = {Intergroup threat theory},
	year = {2017},
	doi = {10.1002/9781118783665.ieicc0162}
	}

@article{golman2017information,
  title={Information avoidance},
  author={Golman, Russell and Hagmann, David and Loewenstein, George},
  journal={Journal of Economic Literature},
  volume={55},
  number={1},
  pages={96--135},
  year={2017},
  doi = {10.1257/jel.20151245}
}

@article{taylor1988illusion,
  title={Illusion and well-being: a social psychological perspective on mental health.},
  author={Taylor, Shelley E and Brown, Jonathon D},
  journal={Psychological bulletin},
  volume={103},
  number={2},
  pages={193},
  year={1988},
  publisher={American Psychological Association}
}

@article{molavi2018,
  title={A theory of non-{Bayesian} social learning},
  author={Molavi, Pooya and Tahbaz-Salehi, Alireza and Jadbabaie, Ali},
  journal={Econometrica},
  volume={86},
  number={2},
  pages={445--490},
  year={2018},
  publisher={Wiley Online Library},
  doi = {10.3982/ECTA14613}
}

@article{banerjee1993economics,
  title={The economics of rumours},
  author={Banerjee, Abhijit V},
  journal={The Review of Economic Studies},
  volume={60},
  number={2},
  pages={309--327},
  year={1993},
  publisher={Wiley-Blackwell}
}

@article{banerjee2004word,
  title={Word-of-mouth learning},
  author={Banerjee, Abhijit and Fudenberg, Drew},
  journal={Games and economic behavior},
  volume={46},
  number={1},
  pages={1--22},
  year={2004},
  publisher={Elsevier}
}

@unpublished{banerjee2020,
  title={Information aggregation under (not so) naive learning},
  author={Banerjee, Abhijit and Olivier Compte},
  note={Mimeo},
  year={2020}
}

@article{compte2004,
  title={Confidence-enhanced performance},
  author={Compte, Olivier and Postlewaite, Andrew},
  journal={American Economic Review},
  volume={94},
  number={5},
  pages={1536--1557},
  year={2004}
}

@article{boxell2018,
  title={A note on internet use and the 2016 US presidential election outcome},
  author={Boxell, Levi and Gentzkow, Matthew and Shapiro, Jesse M},
  journal={Plos one},
  volume={13},
  number={7},
  pages={e0199571},
  year={2018},
  publisher={Public Library of Science San Francisco, CA USA}
}

@article{mian2020,
  title={Coronavirus: the spread of misinformation},
  author={Mian, Areeb and Khan, Shujhat},
  journal={BMC medicine},
  volume={18},
  number={1},
  pages={1--2},
  year={2020},
  publisher={Springer}
}

\appendix
\setcounter{equation}{0}
\renewcommand{\theequation}{A-\arabic{equation}}

\section{Proofs} \label{A-alpha}

\noindent \textbf{Proof of Proposition \ref{uniqueness}.} 
We combine equations \eqref{rho01} and \eqref{rho11} to analyze the law of motion of $\iota^1$:
\begin{equation*}
 \frac{\partial \iota_{t}^{1}}{\partial t} =
\frac{1}{2}(1-\iota^1_{t})\nu k \left[\beta\iota^1_{t}+(1-\beta)\iota^0_{t}\right] - \frac{1}{2}\iota^1_{t}\delta.
\end{equation*}
Define $\vartheta^0 =\beta\iota^0 + (1-\beta)\iota^1$ and $\vartheta^1= \beta\iota^1 + (1-\beta)\iota^0$. Then, in steady state, information prevalence in either group is
\begin{eqnarray}
 \iota^0 & =& \frac{\lambda k\vartheta^0}{1+\lambda k\vartheta^0}, \label{Arho0} \\
 \iota^1 & =& \frac{\lambda k\vartheta^1}{1+\lambda k\vartheta^1}, \label{Arho1}
\end{eqnarray}
from which it follows that there exists one steady state in which $\iota^0=\iota^1=\iota=1-1/(\lambda k)$. To show that this steady state is unique, note that, by \eqref{Arho0} and \eqref{Arho1}, in any positive steady state it must be that

\begin{eqnarray*}
\frac{\iota^0}{\iota^1} &=& \frac{\lambda k \vartheta^0 [1+\lambda k\vartheta^1]}{\lambda k \vartheta^1[1+\lambda k\vartheta^0]}, \\
\iota^0 \vartheta^1[1+\lambda k\vartheta^0 ] &=& \iota^1 \vartheta^0[1+\lambda k\vartheta^1] ,\\
\iota^0 \vartheta^1 - \iota^1 \vartheta^0 &=& \lambda k\vartheta^0\vartheta^1[\iota^1 - \iota^0], \\
\beta \iota^0\iota^1 +(1-\beta)\iota^{0^{2}} - \beta\iota^1\iota^0 - (1-\beta)\iota^{1^{2}} &=& \lambda k\vartheta^0\vartheta^1[\iota^1 - \iota^0], \\
(1-\beta)(\iota^{0^{2}} - \iota^{1^{2}}) &=& \lambda k\vartheta^0\vartheta^1[\iota^1 - \iota^0].
\end{eqnarray*}
If $\iota^1>\iota^0$, then the right-hand side of this equation is positive while the left-hand side is negative, and vice versa for $\iota^1<\iota^0$. It can only hold if $\iota^0=\iota^1=\iota$.

\noindent Finally, either both groups have positive information prevalence, or neither. Deriving the Jacobian of the differential system reveals that both eigenvalues are negative at the positive steady state if and only if $\lambda k>1$. The steady state of zero information prevalence has instead two negative eigenvalues if and only if $\lambda k\leq 1$. These properties of the steady state are inherited from the associated $\rho_0$ and $\rho_1$. This concludes the proof of Proposition \ref{uniqueness}. \hfill $\blacksquare$

\bigskip

\noindent \textbf{Proof of Proposition \ref{exogenous}.} To derive equations \eqref{SSrho0} and \eqref{SSrho1}, note that, by \eqref{rho00}, $\rho^0_0=\iota^0$. By Proposition \ref{uniqueness}, $\iota^1=\iota^0=1-1/(\lambda k)$. Plugging these values in the steady states of equations \eqref{rho01} and \eqref{rho11}, equations \eqref{SSrho0} and \eqref{SSrho1} obtain. 

\noindent The derivatives of \eqref{SSrho0} and \eqref{SSrho1} show that the prevalence of the truth is increasing, while the prevalence of the rumor is decreasing, in $\ell$ and $h$. Hence, the lowest value that $\rho_0$ can take, and the highest value of $\rho_1$, is at $\ell=h=0$, when they are both equal to $\iota$. The truth always has at least as high a prevalence as the rumor. This concludes the proof of Proposition \ref{exogenous}. \hfill $\blacksquare$

\bigskip

\noindent \textbf{Proof of Proposition \ref{Existence}.} \added{Suppose that the prior $y$ is such that it is optimal for individuals who do not verify a message to hold an opinion in line with their bias. Then,} the equilibrium conditions on $\ell , h \in [0,\bar{x}]$ translate into
\begin{align*}
 \added{\frac{c (y (\beta + (1 - \beta) h + \beta (\ell - h)) + (1 - 
 y) \beta (1 - h))}{\beta(1-y)(1-h)}} &\in \Delta ( \ell ), \\
 \added{\frac{c(y(1-\beta)(1-h)+(1-y)(1-\beta+\beta \ell))}{(1-y)(1-\beta +\beta \ell) }}&\in \Delta ( h ).
\end{align*}
\added{As a consequence of Assumption \ref{H-1}, $\Delta (x(\alpha))$ is an upper hemicontinuous, non-empty, closed, and convex correspondence and the values of the left and right derivative of $x$ always exist. Define $G$ as the correspondence from $[0,\bar{x}]^2$ that applies to the functions on the left-hand side of the above two equations a couple $( \ell, h)$ and then applies correspondence $g$ to both solutions. $G$ is a continuous convex valued correspondence from $[0,\bar{x}]^2$ to itself. By Kakutani's fixed point theorem, there exists an equilibrium consisting of $(\ell,h)$ that are a fixed point of $G$.}
\\
\added{For some verification functions, $\bar{x}$ may not be part of the domain, as it is only the asymptotic limit of $x(\alpha)$ as $\alpha$ goes to infinity. In this case, the function $x(\alpha)$ is always strictly increasing and concave. Therefore, $\Delta$ is a strictly decreasing function from $(0,\bar{d}]$ to $[0,\bar{x})$.
By \eqref{equb} and \eqref{equ-b}, we then have that $0\leq \ell<h<1$, and $\Delta(h)>0$, which in turn implies $\Delta(\ell)>0$. These facts imply that a fixed point of $G$ is interior.}
\\
\added{We now derive the values of the initial prior $y$ such that holding an opinion in line with one's bias when a message is not verified is optimal, i.e., that $Pr(b=\Phi|m\neq b)\geq .5$. By \eqref{equ-probmnb}, in steady-state this requirement translates into \eqref{cond-y}.}
\added{The left-hand side of \eqref{cond-y} is continuously increasing in $y \in \left(.5,1 \right)$, is $1$ when $y \rightarrow .5$ and diverges to infinity as $y\rightarrow 1$.
From \eqref{FOC-interior}, as Bayes' rule is continuous and increasing in $y$ and $g(\cdot)$ is continuous and weakly decreasing, $\ell$ and $h$ are continuous and decreasing in $y$.
Therefore, the right-hand side of \eqref{cond-y} is continuous and decreasing in $y$, it is always greater than $1$ and it converges to $1$ as $y\rightarrow 1$, as at that limit both $h$ and $\ell$ are null.
Hence, there exists a unique threshold $\bar{y}>.5$ such that for all $y\geq \bar{y}$ holding an opinion in line with one's bias when a message is not verified is optimal.} This concludes the proof of Proposition \ref{Existence}. \hfill $\blacksquare$

\bigskip


\noindent \textbf{Proof of Proposition \ref{verif}.} The result that in equilibrium, $\alpha^b\leq \alpha^{-b}$ 
follows directly from $Pr(b=\Phi|m=b)\geq Pr(b=\Phi|m\neq b) $. \added{From equations \eqref{equb} and \eqref{equ-b}, we see that neither $k$ nor $\lambda$ affect $\alpha^b$ and $\alpha^{-b}$. We also see that there are two non--negative numbers $L \in \Delta ( \ell )$ and $H \in \Delta ( h ) $ such that }
\begin{align}
 L & = \added{ c \left( \frac{y (\beta + (1-\beta) h - \beta (h- \ell )) }{\beta(1-y)(1-h)} +1 \right)}, \label{Leq} \\
 H & = \added{c \left( \frac{y(1-\beta)(1-h)}{(1-y)(1-\beta +\beta \ell) } +1 \right) }. \label{Heq} 
\end{align}

\noindent \added{To find $\underline{c}$, we need to look for the threshold at which $\ell$ becomes positive, setting $\ell=0$ in the left-hand side of \eqref{Leq}, and solving for this value being equal to $\bar{d}$, which is the derivative of the verification function $x(\cdot)$ at the origin.
Making $c$ explicit, this results in:
}
\[
\underline{c} = \frac{ \bar{d} \beta(1-h)(1-y)}{\beta+h (y-(1+y)\beta)},
\]
\added{which, as a first order effect, is decreasing in $h$ and $y$, and increasing in $\beta$. However, from equation \eqref{equ-b}, when $\ell=0$ we have}
\begin{eqnarray}\label{ttrr_l0}
\added{h = g \left( c \left( 1+ \frac{y (1-h)}{1-y} \right) \right).} 
 \end{eqnarray}
\added{Therefore, $h$ is independent of $\beta$ and increasing in $y$, and then $\underline{c}$ is decreasing in $y$ and increasing in $\beta$.}
\noindent \added{To find $\bar{c}$, we look for the threshold at which $h$ becomes positive, setting $\ell=0$ and $h=0$ in the left-hand side of \eqref{Heq}, and solving for this value being equal to $\bar{d}$. In this way, making $c$ explicit, we find $ \bar{c} = \bar{d}(1-y)$.}

\noindent Finally, as $\lambda$ and $k$ do not affect verification rates, stability follows from Proposition \ref{uniqueness}. This concludes the proof of Proposition \ref{verif}. \hfill $\blacksquare$

\bigskip






\bigskip

\noindent \textbf{Proof of Proposition \ref{ttrr}.} 
\added{We start from the computations in the proof of Proposition \ref{verif}. Expressing equation \eqref{Leq} in terms of $\ell$, we obtain}
\begin{align*}
 \ell & = \added{h-\frac{1-h}{y}}+\frac{(1-h)(1-y)L}{c y}-\frac{h}{\beta}. 
\end{align*}
Plugging this into the truth to rumor ratio from \eqref{truthtorumor}, we obtain
\begin{equation*}
\frac{\rho_0}{\rho_1} = \added{1+ \frac{2}{y} \left(\frac{L(1-y)}{c} -1 \right) \beta}.
\end{equation*}
The results follow from the fact that if $L\in \Delta(\ell)$, then $g(L) = \ell$. Finally, note that from \eqref{truthtorumor} the truth to rumor ratio when $\alpha^b=0$, i.e., \added{$l=0$}, is
\begin{equation*}
\label{eq_ttrr}
 \frac{\rho_0}{\rho_1} = \frac{1+h - 2\beta h}{1-h}.
\end{equation*}
\added{From equation \eqref{equ-b}, when $\ell=0$ we have \eqref{ttrr_l0}. This shows that $h$ is independent of $\beta$ and $\rho_0/\rho_1$ is decreasing in homophily $\beta$ if $\alpha^b$ stays at $0$.}
This concludes the proof of Proposition \ref{ttrr}. \hfill $\blacksquare$

\bigskip

\noindent \textbf{Proof of Proposition \ref{partisan}.} As individuals of type $0$ are always of the opinion that $\Phi=0$, $\gamma$ is irrelevant for $\rho_{0,t}$. We separately consider partisans and non-partisans of type $1$. All \added{informed} partisans of this type hold opinion $\Phi=1$; denote the corresponding prevalence as $\rho^\gamma_{1,t}$. We denote the prevalence of opinion $b$ among non-partisans of type $1$ as $\rho^{1-\gamma}_{b,t}$ and the proportion of informed non-partisans (partisans) as $\iota^{1-\gamma}_{t}$ ($\iota^{\gamma}_{t}$). As in the benchmark model, $\iota^0_t=\rho^0_{0,t}$, but now $\iota^1_t=\gamma \iota^\gamma_t+ (1-\gamma)[\iota^{1-\gamma}_{0,t}+\iota^{1-\gamma}_{1,t}]$. Hence,
\begin{align*}
\rho_{1,t} &= \frac{1}{2}[(1-\gamma)\rho^{1-\gamma}_{1,t}+\gamma \rho^{\gamma}_{1,t}] 
\text{ and } \rho_{0,t} = \frac{1}{2} [\iota^0_t + (1-\gamma)\rho^{1-\gamma}_{0,t}].
\end{align*}
The system describing the evolution of the prevalence of opinions is
\begin{align}
 \frac{\partial \rho^0_{0,t}}{\partial t} & = \frac{1}{2} (1-\rho^0_{0,t})\nu k [\beta\iota^0_t +(1-\beta)\iota^1_t] - \frac{1}{2} \rho^0_{0,t}\delta, \label{stubborn0} \\
 \frac{\partial \rho^\gamma_{1,t}}{\partial t} & = \frac{1}{2}\gamma(1-\rho^\gamma_{1,t})\nu k[\beta\iota^1_t +(1-\beta)\iota^0_t] - \frac{1}{2}\gamma \rho^\gamma_{1,t}\delta, \label{stubborn1} \\ 
 \frac{\partial \rho^{1-\gamma}_{0,t}}{\partial t} & = \frac{1}{2} (1-\gamma) [1-\rho^{1-\gamma}_{0,t}-\rho^{1-\gamma}_{1,t}] \nu k \left[ \beta \ell[(1-\gamma)\rho^{1-\gamma}_{1,t}+\gamma \rho^{\gamma}_{1,t}]+ \right. \nonumber \\ 
 & \left. + \beta h (1-\gamma)\rho^{1-\gamma}_{0,t} +(1-\beta)h \rho^0_{0,t}\right] - \frac{1}{2} (1-\gamma) \rho^{1-\gamma}_{0,t}\delta , \label{stubborn0g} \\
 \frac{\partial \rho^{1-\gamma}_{1,t}}{\partial t} & = \frac{1}{2} (1-\gamma) [1-\rho^{1-\gamma}_{0,t}-\rho^{1-\gamma}_{1,t}] \nu k \left[ \beta (1- \ell)[(1-\gamma)\rho^{1-\gamma}_{1,t}+\gamma \rho^{\gamma}_{1,t}] + \right. \nonumber \\ 
 & \left. + (1- h) \left( \beta (1-\gamma)\rho^{1-\gamma}_{0,t} +(1-\beta) \rho^0_{0,t} \right) \right] - \frac{1}{2} (1-\gamma) \rho^{1-\gamma}_{1,t}\delta. \label{stubborn1g}
\end{align}
By combining equations \eqref{stubborn0g} and \eqref{stubborn1g}, we find that the evolution of $\iota^{1-\gamma}_t=\rho^{1-\gamma}_{0,t}+\rho^{1-\gamma}_{1,t}$ mirrors the one of $\iota^\gamma_t$. 
Following the same analysis of the benchmark model, we find that the steady state values of $\rho_0$ and $\rho_1$ are
\begin{align*}
\rho_0 &= \frac{1}{2}\frac{1+(1-\gamma)h+2\beta(1-\gamma)(\ell-h)}{1+\beta(1-\gamma)(\ell-h)}\iota 
\text{ and } \rho_1=\frac{1}{2}\frac{1-(1-\gamma)h}{1+\beta(1-\gamma)(\ell-h)}\iota.
\end{align*}
Hence, verification with partisans is as in the benchmark model with $\hat{x}(\alpha)=(1-\gamma)x(\alpha)$. The prevalence of both opinions and verification rates are therefore unchanged. This concludes the proof of Proposition \ref{partisan}. \hfill $\blacksquare$

\section{Computations for Examples}
\label{app:examples}

\noindent \textbf{Computations for Example \ref{ex:exp_with_cap}.} \added{Here, we compute the different possible equilibria with the exponential verification function with a cap at $\bar{x}$.
\\
\textit{Case I.} By Proposition \ref{verif} and since $\bar{d}=1$ in this example, no verification happens, i.e., $\ell_1=h_1=0$, if $c\geq 1-y$. Therefore, in all the following cases with verification, $c< 1-y$ must hold.
\\
\noindent \textit{Case II.} If there is some verification, there are several possible equilibria. We first focus on cases where only one message is verified, i.e., $\ell_2=0$. We look for $h$ that solves \eqref{Heq}, setting $H(h_2)=1-h_2$. The solution is}
\begin{eqnarray*}
\added{h_2= \frac{1-y-c}{1-y-cy},}
\end{eqnarray*}
\added{which is always non negative if $c<1-y$. Moreover, $h_2<\bar{x}$ if
\[
c >\frac{(1-\bar{x})(1-y)}{1-\bar{x} y}=c_1
\]
To derive when $\ell_2=0$, we set the left-hand side of \eqref{Leq} lower than 1, and obtain that this requires $y\geq\beta$.
In this case, it is possible to compute explicitly $\bar{y}_2= 1/(1+2c)$.}
\\
\noindent \textit{Case III.} \added{If $c\leq c_1$, we have an equilibrium with $\ell_3=0$ and $h_3=\bar{x}$; again, $\ell_3=0$ is guaranteed if $y\geq \beta$. The condition on priors requires $y\geq \bar{y}_3= 1/(2-\bar{x})$. Therefore, this constraint is binding only when $\beta<1/(2-\bar{x})$.}
\\
\noindent \textit{Case IV.} \added{In the equilibrium where $0<\ell,h,<\bar{x}$, the system from equations \eqref{Leq} and \eqref{Heq} can be solved substituting $L$ with $1-\ell$ and $H$ with $1-h$. If $\ell,h>0$, this results in}
\[
\added{\ell_4 = \frac{1-c-y}{1-y} \cdot \frac{\beta-y}{\beta}} \text{ and }
\added{h_4 = 
 \frac{1-c-y}{1-y} \cdot \frac{1-(1-c)y-c\beta}{1-y-c \beta}.}
\]
\added{Hence, $\ell_4>0$ needs $\beta>y$, and this also implies $h_4>0$ as $h_4>\ell_4$; moreover, $h_4<\bar{x}$ holds if the left-hand side of \eqref{Heq} is lower than $1-\bar{x}$, which translates into
\[
c > \frac{\beta (1-\bar{x}) (1-y)}{\beta -\bar{x} (\beta -(1-\beta ) y)}=c_2
\]
Note that both $\ell_4$ and $h_4$ are decreasing in $c$ and increasing in $\beta$ for $\beta \in \left(.5, 1 \right)$ and $y \in \left(.5, 1 \right)$. In this equilibrium, \eqref{cond-y} holds if
\[
y\geq \bar{y}_4=\frac{1}{2} \left(2+c -\sqrt{c} \sqrt{4+c-4 \beta +4 \beta ^2 c-4 \beta c}-2 \beta c\right).
\]
\noindent \textit{Case V.} Let us consider now the equilibrium in which $0<\ell_5< \bar{x}$ and $h_5=\bar{x}$. First, we set the right-hand side of \eqref{Leq} to $1-l$, and we solve for $\ell$, obtaining 
\[
\ell_5 = 1-c\frac{(1-\beta ) \bar{x} y+\beta (1-\bar{x}+y)}{\beta (c y+(1-\bar{x}) (1-y))}.
\]
which is decreasing in $c$ and increasing in $\beta$. Note that again $\ell_5$ is positive if $y<\beta$. 
To see that $h=\bar{x}$ is possible, we set the right-hand side of \eqref{Heq} weakly smaller than $1-\bar{x}$, and we solve for $c$, obtaining}
\begin{eqnarray*}
\added{c \leq c_3} &=\added{ \frac{1-y}{2}+ \frac{(1-y) \left(1- \bar{x}\beta -\sqrt{(y+1-\beta (1-\bar{x}))^2-4 \bar{x} y}\right)}{2
 (\beta -y)}}&
\end{eqnarray*}
\added{This is an equilibrium only when $\ell_5<\bar{x}$; this holds if $1-\ell_5<1-\bar{x}$, that is:
\[
c > \frac{\beta (1-\bar{x})^2 (1-y)}{\beta -\beta \bar{x}+\bar{x} y}=c_4.
\]
The condition \eqref{cond-y} on the prior is satisfied if}
\begin{eqnarray*}
 \added{y\geq \bar{y}_5} & \added{= \frac{c-\beta \bar{x}+\bar{x}-3+\beta +\sqrt{(3-\beta -c+\beta \bar{x}-\bar{x})^2+4 (1-\beta c) (\beta -2 \beta c+2 c-\beta \bar{x}+\bar{x}-2)}}{2 (\beta -2 \beta c+2 c-\beta \bar{x}+\bar{x}-2)}.}&
\end{eqnarray*}
\added{Note that this condition implies $c>c_4$.}
\\
\noindent \textit{Case VI.} \added{Finally, there is an equilibrium with $\ell_6=h_6=\bar{x}$ if $c\leq c_4$. In this case, condition \eqref{cond-y} on the prior is satisfied if}
\[
\added{y\geq \bar{y}_6= \frac{1-\beta +\beta \bar{x}}{2 (1-\beta ) (1-\bar{x})+\bar{x}}.}
\]
\added{When $\bar{x}\rightarrow 1$, $\ell,h\rightarrow 1$, and this is an equilibrium only for $c=0$.}
\\
\added{Summing up this example, $\ell_1=h_1=0$ if $c\geq 1-y$; if instead $c<1-y$, there are the following equilibria:}
\begin{enumerate}[label=\Alph*),leftmargin=*]
\item \textit{i)} \added{$\ell_1=h_1=0$ if $c\geq 1-y$;}
\\
\item \added{if $c<1-y$ and $y\geq \beta$, $\ell=0$, the following cases emerge:}
\\
\textit{ii)} \added{$\ell_2=0$ and $h=h_2\in (0,\bar{x})$ if $c_1<c$ and $y\geq \bar{y}_1$;}
\\
\textit{iii)} \added{$\ell_3=0$ and $h=h_3=\bar{x}$ if $c\leq c_1$ and, whenever $\beta<\bar{y}_2$, $y\geq \bar{y}_2$;}
\item \added{if $c<1-y$ and $y<\beta$, $\ell>0$, the following cases emerge:}
\\
\textit{iv)} \added{$0<\ell_4<h_4<\bar{x}$ if $c>c_2$ and $y\geq \bar{y}_4$;}
\\
\textit{v)} \added{$\ell_5\in (0,\bar{x})$ and $h_5=\bar{x}$ if $c_3\leq c < c_4$ and $y\geq \bar{y}_5$;}
\\
\textit{vi)} \added{$\ell_6=h_6=\bar{x}$ if $c\leq c_4$ and $y\geq \bar{y}_6$.}
\end{enumerate}
\added{Note that not all equilibria necessarily exist for all parameters values. For example, when $\bar{x}\rightarrow 1$, we have an exponential verification function without any cap, and we only have three possible equilibria:}
\\
\textit{i)} \added{$\ell_1=h_1=0$ if $c\geq 1-y$;}
\\
\textit{ii)} \added{$\ell_2=0$ and $h_2\in (0,1)$ if $c<1-y$, $y\geq \beta$ and $y\geq\bar{y}_2$;}
\\
\textit{iii)} \added{$0<\ell_4<h_4<\bar{x}$ if $c<1-y$, $y< \beta$ and $y\geq\bar{y}_4$.}
\\
\added{This example is depicted in Figure \ref{fig:exp}.} \hfill $\blacksquare$

\bigskip

\noindent \textbf{Computations for Example \ref{ex:marginal_homphily}.}
From equation \eqref{truthtorumor} we have the expression of the truth to rumor ratio, as a function of $\ell$ and $h$.
If we are in the corner solution where $\ell=0$ and $h=0$, then the truth to rumor ratio is one and remains so if $\beta$ changes marginally.
\\
\added{For all the other cases discussed in the example, we plug the values of $\ell$ and $h$, as obtained in Example \ref{ex:exp_with_cap}, into equation \eqref{truthtorumor}. In this way we obtain that the truth to rumor ratio is constant in homophily when $0<\ell<h<0$ and when $\ell=h=\bar{x}$.}
\\
\added{The truth to rumor ratio when $\ell\in(0,\bar{x})$ and $h=\bar{x}$ is}
\begin{eqnarray*}
\added{\frac{\rho_0}{\rho_1}=\frac{(1+\bar{x}) (1-y)+c y + 2\beta ((1-y)(1-\bar{x})-c)}{(1-\bar{x}) (1-y)+c y},}
\end{eqnarray*}
\added{which is increasing in $\beta$ if 
$y\leq (1-c-\bar{x})/(1-\bar{x})$, and decreasing otherwise. Hence, the effect of homophily on the truth to rumor ratio is positive if $y\in[y_2,(1-c-\bar{x})/(1-\bar{x}) 
]$ and negative if $y\in((1-c-\bar{x})/(1-\bar{x})
,1]$.} \hfill $\blacksquare$

\end{document}